\documentclass[12pt,a4paper]{article}
\pdfoutput=1
\usepackage{ifthen} 
\newboolean{pdflatex}
\usepackage[normalem]{ulem}
\setboolean{pdflatex}{true} 

\newboolean{articletitles}
\setboolean{articletitles}{true} 
\newboolean{uprightparticles}
\setboolean{uprightparticles}{false} 
\usepackage{multirow}
\usepackage[dvipsnames]{xcolor}
\usepackage{savesym}
\savesymbol{checkmark}
\usepackage{pifont}
\usepackage{braket}
\usepackage[dvipsnames]{xcolor}
\usepackage[pdftitle={Kolya: an open-source package for inclusive semileptonic B decays},
  pdfauthor={Matteo Fael, Ilija Milutin, Keri Vos},
  pdfkeywords={},
  bookmarks=true, linktocpage,
  colorlinks=true, allbordercolors=white, allcolors=blue]{hyperref}

\usepackage{mciteplus}
\usepackage[numbers,sort&compress]{natbib}
\usepackage{amsmath}
\usepackage{amssymb}
\usepackage{graphicx}
\usepackage[frozencache,cachedir=.]{minted}
\usepackage[colorlinks=true, allcolors=blue]{hyperref}

\usepackage{listings}
\usepackage{xcolor}
\usepackage{comment}
\definecolor{codegreen}{rgb}{0,0.6,0}
\definecolor{codegray}{rgb}{0.5,0.5,0.5}
\definecolor{codepurple}{rgb}{0.58,0,0.82}
\definecolor{backcolour}{rgb}{0.95,0.95,0.92}

\lstdefinestyle{mystyle}{
    backgroundcolor=\color{backcolour},   
    commentstyle=\color{codegreen},
    keywordstyle=\color{magenta},
    numberstyle=\tiny\color{codegray},
    stringstyle=\color{codepurple},
    basicstyle=\ttfamily\footnotesize,
    breakatwhitespace=false,         
    breaklines=true,                 
    captionpos=b,                    
    keepspaces=true,                 
    numbers=left,                    
    numbersep=5pt,                  
    showspaces=false,                
    showstringspaces=false,
    showtabs=false,                  
    tabsize=2
}

\begin{document}
\begin{titlepage}

\begin{flushright}
{\small
Nikhef 2023-022\\
CERN-TH-2024-155\\
SI-HEP-2024-20\\
P3H-24-065\\
\today \\
}
\end{flushright}

\vskip1cm
\begin{center}
{\Large \bf\boldmath Kolya: an open-source package for inclusive semileptonic $B$ decays}
\end{center}

\vspace{0.5cm}
\begin{center}
{\sc Matteo Fael$^{a}$, Ilija S.\ Milutin$^{b}$, and K. Keri Vos$^{c,d}$} \\[3mm]

{\it $^a$ Theoretical Physics Department, CERN,\\
1211 Geneva, Switzerland}\\[0.3cm]

{\it $^b$ Theoretische Physik 1, Center for Particle Physics Siegen \\
   Universit\"at Siegen,  D-57068 Siegen, Germany}\\[0.3cm]

{\it $^c$Gravitational 
Waves and Fundamental Physics (GWFP),\\ 
Maastricht University, Duboisdomein 30,\\ 
NL-6229 GT Maastricht, the
Netherlands}\\[0.3cm]

{\it $^d$Nikhef, Science Park 105,\\ 
NL-1098 XG Amsterdam, the Netherlands}
\end{center}

\vspace{0.6cm}
\begin{abstract}

\vskip0.2cm\noindent
We introduce the code \texttt{kolya}, an open-source tool for phenomenological analyses of inclusive semileptonic $B$ meson decays. 
It contains a library to compute predictions for the total rate and 
various kinematic moments within the framework of the heavy quark expansion,
utilizing the so-called kinetic scheme.
The library currently includes power corrections up to $1/m_b^5$.
All available QCD perturbative corrections are implemented 
via interpolation grids for fast numerical evaluation.
We also include effects from new physics parameterised as Wilson coefficients
of dimension-six operators in the weak effective theory below the electroweak scale.
The library is interfaced to CRunDec for easy evaluation of the quark masses
and strong coupling constant at different renormalization scales.
The library is developed in Python and does not require compilation.
It can be used in an interactive Jupyter notebook session.
\end{abstract}

\end{titlepage}


\section{Introduction}
Measurements of semileptonic $B$ decays lie at the core
of the Belle II and LHCb physics program in the upcoming years. 
Thanks to relatively large rates and clean experimental signatures, inclusive
and exclusive semileptonic decays with a $b \to c l \bar \nu_l$  transition $(\ell = e, \mu)$ offer a clean avenue for the
determinations of $|V_{cb}|$, the element of the Cabibbo-Kobayashi-Maskawa matrix (CKM) which parameterizes the strength of the weak interaction among bottom and charm quarks in the Standard Model of particle physics.

Inclusive determinations of $|V_{cb}|$ exploit that 
the semileptonic rate $\Gamma_\mathrm{sl}$ and moments of kinematic spectra can be described with good precision using the heavy quark expansion (HQE)~\cite{Manohar:1993qn,Blok:1993va,Bigi:1993fe}.
In the HQE, these observables are expressed as a series of non-perturbative HQE elements proportional 
to increasing powers of the inverse bottom quark mass times the QCD scale parameter, 
$\Lambda_\mathrm{QCD}/m_b$. In addition, each order in the HQE also receives corrections expressed as a series expansion in the
strong coupling constant, $\alpha_s$, which can be systematically
calculated in perturbative QCD.

This paper documents the first release of the open-source code \texttt{kolya}~\cite{fael_2024_10818195}. It consists of a Python library which computes the prediction
for the total rate and lepton-energy, hadronic invariant mass and dilepton invariant mass kinematic moments within 
the framework of the HQE utilizing the so-called kinetic scheme~\cite{Bigi:1996si,Czarnecki:1997sz,Fael:2020iea,Fael:2020njb}. The \texttt{kolya} code supersedes and extends the 
code developed for the fit of $q^2$ moments~\cite{Bernlochner:2022ucr} measured by Belle~\cite{Belle:2021idw} and Belle~II~\cite{Belle-II:2022evt}.
We also include effects from new physics (NP) studied in Ref.~\cite{Fael:2022wfc}. These are parameterised as Wilson coefficients of dimension-six operators in
the weak effective theory below the electroweak scale~\cite{Aebischer:2017gaw,Jenkins:2017jig}.

Several building blocks necessary for the prediction of $\Gamma_\mathrm{sl}$ 
and the moments to high orders in $\alpha_s$ and  the $1/m_b$ expansion have been presented over the last 30 years  (see
Sec.~\ref{subsec:implementation} for an exhaustive list of references).
The \texttt{kolya} library provides the first comprehensive open-source framework in which all available
corrections are implemented and validated.
A schematic overview of the perturbative corrections implemented for the total 
rate and the moments is given in Tab.~\ref{tab:overview}.
This document accompanies the first release of \texttt{kolya} and details the specifics of the code.
Although this paper represents a reference for future analyses of Belle II measurements of inclusive $B \to X_c l \bar \nu_l$ decays and gives 
a first outlook of \texttt{kolya} with
basic examples to try in a Jupyter notebook, it is
not meant to be a review article on semileptonic decays. 
To obtain a deeper understanding of the scientific 
part, the user is referred to e.g.\ Refs.~\cite{Gambino:2013rza,Dingfelder:2016twb,Fael:2024rys}.
The software \texttt{kolya} complements in scope
several other open-source packages in HEP, 
in particular \texttt{flavio}~\cite{Straub:2018kue}, 
\texttt{EOS}~\cite{EOSAuthors:2021xpv}, \texttt{HEPfit}~\cite{DeBlas:2019ehy}, \texttt{HAMMER}~\cite{Bernlochner:2020tfi} and \texttt{SuperIso}~\cite{Arbey:2011zz}.

This article is structured as follows. 
In Sec.~\ref{sen:definitions}, we present the definitions of observables. Their implementation in the code is discussed in Sec.~\ref{sec:implementation}
where we discuss various ingredients implemented
and quote the original references from which the material was obtained. 
The definition of the effective Hamiltonian parametrizing 
NP effects is given in Sec.~\ref{sec:NP}.
Section~\ref{sec:usage} focuses on basic usage of the code, 
illustrating the installation, the classes implemented in the code,
use of the code for calculating $\Gamma_\mathrm{sl}$ and the moments
together with details about our validation of the code.
We close in Sec.~\ref{sec:conc} with an outlook.

\section{Definitions}
\label{sen:definitions}
We consider the semileptonic decay 
\begin{equation}
        \overline B(p_B) \to X_c(p_X) l(p_l) \bar \nu_l(p_\nu) \text{ with } l = e, \mu,
\end{equation}
and describe the decay rate in the rest frame of the $B$ meson, i.e.\ $p_B = (M_B,\vec 0)$.
The leptons are considered massless. 
We denote the total momentum of the lepton pair by $q = p_l + p_\nu$, the total momentum of the hadronic system by $p_X=p_B -q$, and the electron energy by $E_l$.

Within the HQE, it is possible to make a prediction for the various differential rates w.r.t.\ 
the $q^2$, the total leptonic energy $q_0 = E_\nu + E_l$, and the energy $E_l$ of the charged lepton.
However, the predictions for the differential rates cannot be compared point by point with data. This is because, on the one hand, the phase space region allowed at the parton level is smaller than the physical one.
On the other hand, power corrections become 
singular close to the endpoint.
Instead, data of inclusive $\overline{B}\to X_c l \bar{\nu}_l$ has to be compared to theory predictions of
integrated quantities like the total rate 
\begin{equation}
    \Gamma_\mathrm{sl} = 
    \int \frac{\text{d}^3 \Gamma}{\text{d}q^2 \, \text{d}q_0 \, \text{d}E_l}
    \text{d}q^2 \, \text{d}q_0 \, \text{d}E_l
    \label{eqn:GammaSL}
\end{equation}
or moments of the differential distribution of some observable $O$, where $O = E_l, q^2, M_X^2$, normalized w.r.t.\ the partial decay width.
The moments are defined by
\begin{equation}
    \langle (O)^n \rangle_{\mathrm{cut}} =
    \int_{\mathrm{cut}}
    (O)^n 
    \frac{\mathrm{d} \Gamma}{\mathrm{d} O} \, \mathrm{d} O
    \Bigg/
    \int_{\mathrm{cut}}
    \frac{\mathrm{d} \Gamma}{\mathrm{d} O} \, \mathrm{d} O\ ,
    \label{eqn:defOmoments}
\end{equation}
where $\mathrm{d} \Gamma / \mathrm{d} O$ is the 
differential rate for the variable $O$.
The subscript ``cut'' generically denotes some restriction in the lower integration limit. 
From the theoretical side, the dependence of the moments on a lower cut yields additional information on the HQE parameters and thus provides
a better handle for their extraction via global fits.
From the experimental side, the spectrum is usually not measurable
entirely due to detector acceptance. For example at the $B$-factories, 
a lower cut on the charged lepton energy, $E_\ell \ge E_\mathrm{cut}$ with $E_\mathrm{cut} \simeq 0.5$ GeV, is applied
to suppress the background.
It is also possible to consider the partial decay width with a cut on $E_l$ or $q^2$, defined by
\begin{align}
    \Delta \Gamma_\mathrm{sl} (E_\mathrm{cut}) &= 
    \int_{E_l \ge E_\mathrm{cut}} 
    \frac{\mathrm{d} \Gamma}{ \mathrm{d} E_l }\, 
    \mathrm{d} E_l\ , &
    \Delta \Gamma_\mathrm{sl} (q^2_\mathrm{cut}) &= 
    \int_{q^2 \ge q^2_\mathrm{cut}} 
    \frac{\mathrm{d} \Gamma}{ \mathrm{d} q^2 }\, 
    \mathrm{d} q^2.
\end{align}

For higher moments ($n \ge 2$), usually centralized moments are considered. The centralized moments of the charged lepton energy are then defined as:
\begin{align}
    \ell_1 (E_\mathrm{cut}) &= \langle E_l \rangle_{E_l \ge E_\mathrm{cut}}\ , & 
    \ell_n (E_\mathrm{cut}) &= \Big\langle (E_l - \langle E_l\rangle )^n \Big\rangle_{E_l \ge E_\mathrm{cut}} 
    \text{ for } n\ge2\ ,
    \label{eqn:defellcentralized}
\end{align}
and moments of the hadronic invariant mass as:
\begin{align}
    h_1(E_\mathrm{cut}) &= \langle M_X^2 \rangle_{E_l \ge E_\mathrm{cut}}\ , & 
    h_n (E_\mathrm{cut}) &= \Big\langle (M_X^2 - \langle M_X^2 \rangle )^n \Big\rangle_{E_l \ge E_\mathrm{cut}}
    \text{ for } n\ge2\ .
    \label{eqn:defhicentralized}
\end{align}
The first moments with $n=1$ correspond to the mean value of observable over the considered
integration domain, while the second centralized moments are the variances of the distributions. Experimentally, moments up to $n=3$ and $n=4$ are currently available. 
Ref.~\cite{Fael:2018vsp} proposed also the study of the moments of the $q^2$ spectrum, defined by
\begin{align}
    q_1(q^2_\mathrm{cut}) &= \langle q^2 \rangle_{q^2 \ge q^2_\mathrm{cut}}\ , & 
    q_n (q^2_\mathrm{cut}) &= \Big\langle (q^2 - \langle q^2 \rangle )^n \Big\rangle_{q^2 \ge q^2_\mathrm{cut}}
    \text{ for } n\ge2\ .
    \label{eqn:defqicentralized}
\end{align}
These $q^2$ moments are invariant under reparametrization and therefore they depend on a reduced set of 
HQE parameters (which we will introduce in Sec.~\ref{sec:implementation}), like the total rate (see Refs.~\cite{Manohar:2010sf,Fael:2018vsp}). 
Since a cut on $E_l$ in the definition of the $q^2$ moments would break reparametrization invariance (RPI),
Ref.~\cite{Fael:2018vsp} suggested to consider instead a lower cut on $q^2$. Such a lower cut on $q^2$ also imposes an indirect cut on the charged lepton energy through
\begin{equation}
    E_l \ge \frac{M_B^2+q^2_\mathrm{cut}-M_D^2 - \lambda^{1/2}(M_B^2,q^2_\mathrm{cut}, M_D^2)}{2M_B}\ ,
\end{equation}
where $\lambda(a,b,c) = a^2+b^2+c^2-2ab-2ac-2bc$ is the K\"all\'en function.
A cut on $q^2$ is therefore capable of excluding low-energy electrons in the 
experimental analysis, on equal footing as a cut on $E_l$.

\section{Implementation in the SM}
\label{sec:implementation}
\subsection{Building blocks}
We use the heavy quark expansion (HQE) and express the total semileptonic width $\Gamma_\mathrm{sl}$ 
and the kinematic moments as a double expansion in $1/m_b$ and $\alpha_s$. While working with the HQE, it is often advantageous to consider dimensionless quantities normalized w.r.t.\ the bottom quark mass $m_b$.
We will denoted them with a ``hat($\hat{\phantom a}$)'': e.g.\ $\hat q^2 = q^2/m_b^2, \hat E_l = E_l /m_b$.

As a starting point, we define the following building blocks:
\begin{align}\label{eq:Qijbuild}
    Q_{ij} =& \frac{1}{\Gamma_0}
    \int_\mathrm{cut} 
    \text{d}E_l \, \text{d}q_0 \, \text{d}q^2 \, 
    (q^2)^i 
    (q_0)^j
    \, 
    \frac{\text{d}^3\Gamma}{\text{d}q^2 \, \text{d}q_0 \, \text{d}E_l} \ ,
    \end{align}
   where $q_0 = (E_l + E_\nu) = v \cdot q$ is the total leptonic energy  with $v = p_B/M_B$ and $q^2$ is the leptonic invariant mass. 
   Schematically, we write\begin{align}
    Q_{ij} =& 
    (m_b)^{2i+j} 
    \Bigg[ 
    Q_{i,j}^{(0)} + 
    Q_{i,j}^{(1)} \frac{\alpha_s(\mu_s)}{\pi} + 
    Q_{i,j}^{(2)} \left( \frac{\alpha_s(\mu_s)}{\pi} \right)^2+
    \frac{\mu_\pi^2}{m_b^2}
    \Bigg(
        Q_{i,j, \pi}^{(0)} + 
        Q_{i,j, \pi}^{(1)} \frac{\alpha_s(\mu_s)}{\pi} 
    \Bigg)
    \notag \\ &
    \frac{\mu_G^2(\mu_b)}{m_b^2}
    \Bigg(
        Q_{i,j, G}^{(0)} + 
        Q_{i,j, G}^{(1)} \frac{\alpha_s(\mu_s)}{\pi} 
    \Bigg)
    +\frac{\rho_D^3(\mu_b)}{m_b^3}
    \Bigg(
        Q_{i,j, D}^{(0)} + 
        Q_{i,j, D}^{(1)} \frac{\alpha_s(\mu_s)}{\pi} 
    \Bigg)
    \notag \\ &
    +\frac{\rho_{LS}^3(\mu_{b})}{m_b^3}
    \Bigg(
        Q_{i,j, LS}^{(0)} + 
        Q_{i,j, LS}^{(1)} \frac{\alpha_s(\mu_s)}{\pi} 
    \Bigg)
    +O\left(\frac{1}{m_b^4}\right)
    \Bigg]\ ,
    \label{eqn:defQij}
\end{align}
where
\begin{equation}
    \Gamma_0 = 
    \frac{m_b^5 G_F^2 A_\mathrm{ew}}{192 \pi^3} | V_{cb}|^2\ .
\end{equation}
The factor $A_\mathrm{ew} = 1.01435$ stems from short-distance radiative 
corrections at the electroweak scale~\cite{Sirlin:1977sv}. 
$\alpha_s \equiv \alpha_s^{(n_f)}(\mu_s)$ is
the strong coupling constant taken with $n_f$ active quarks and at the renormalization scale $\mu_s$.
To leading order in $1/m_b$, the heavy $B$ meson decay coincides with the decay of a free bottom
computed in perturbative QCD. Starting from $O(1/m_b^2)$ the predictions depend
on a set of HQE parameters: non-perturbative matrix elements of local operators. 
These are denoted by $\mu_\pi^2, \mu_G^2, \rho_D^3,\rho_{LS}^3$. 
The tree-level 
expressions are known also to higher orders in $1/m_b$ (see Refs.~\cite{Dassinger:2006md,Mannel:2010wj,Fael:2018vsp,Mannel:2023yqf}).
They are implemented in \texttt{kolya} up to $1/m_b^5$.
However, in \eqref{eqn:defQij} they are omitted to keep 
a compact notation.
The explicit definitions of the HQE parameters up to $1/m_b^5$ are reported in Appendix~\ref{app:HQE_def}. The HQE parameters in \eqref{eqn:defQij} are quoted in what we refer to as the ``historical'' basis employed in e.g. \cite{Bordone:2021oof,Finauri:2023kte}. For RPI quantities, like $q^2$ moments, it is however useful to work in the RPI basis, which has a reduced number of parameters \cite{Mannel:2018mqv,Fael:2018vsp}. The differences between these two bases are detailed in Ref.\cite{Mannel:2018mqv,Mannel:2024crj}. 

The functions denoted by $Q_{ij}$ are the fundamental building blocks necessary to assemble the predictions
for the centralized moments $q_i$ in \eqref{eqn:defqicentralized} and $h_i$ in \eqref{eqn:defhicentralized}.
They all depend on the mass ratio
\begin{equation}
    \rho \equiv \frac{m_c}{m_b}\ ,
\end{equation}
where $m_c$ and $m_b$ refer to the on-shell masses of the charm and bottom quark.
In \eqref{eq:Qijbuild}, the subscript ``cut'' refers to certain restrictions in the phase-space integration.
For the prediction of $q_i$ in \eqref{eqn:defqicentralized}, we apply the cut $q^2 > q^2_\mathrm{cut}$
so that various build blocks in \eqref{eqn:defQij} depend on $\rho$ and $\hat q^2_\mathrm{cut}$: 
$Q_{ij} \to Q_{ij} (\rho, \hat q^2_\mathrm{cut})$.
For the hadronic moments $h_n$ in \eqref{eqn:defhicentralized},
the restriction is on the electron energy $E_l > E_\mathrm{cut}$, so that the
building blocks are functions of $\rho$ and $\hat E_\mathrm{cut}$: $Q_{ij} \to Q_{ij} (\rho, \hat E_{\rm{cut}})$ 
(see Eq.~\eqref{eqn:MnDefinition} for the relation between $h_n$ and $Q_{ij}$).

The QCD corrections depend also on the renormalization scale $\mu_s$ of the strong coupling constant
starting at $O(\alpha_s^2)$.
The functions $Q_{i,j,G}, Q_{i,j,D}$ and $Q_{i,j,LS}$ depend on $\mu_b$, the scale at which the Wilson coefficients of the HQET Lagrangian are matched onto QCD, starting 
from $O(\alpha_s)$.

To construct the centralized electron energy moments in \eqref{eqn:defellcentralized},
we consider the moments of the charged-lepton energy $E_l = p_l \cdot v$ within the HQE:
\begin{align}
    L_{i} =& \frac{1}{\Gamma_0}
    \int_{E_l \ge E_\mathrm{cut}} 
    \text{d}E_l \, \text{d}q_0 \, \text{d}q^2 \, 
    (E_l)^i 
    \, 
    \frac{\text{d}^3\Gamma}{\text{d}q^2 \, \text{d}q_0 \, \text{d}E_l}
    \notag \\
    =& 
    (m_b)^{i} 
    \Bigg[ 
    L_{i}^{(0)} + 
    L_{i}^{(1)} \frac{\alpha_s(\mu_s)}{\pi} + 
    L_{i}^{(2)} \left( \frac{\alpha_s(\mu_s)}{\pi} \right)^2+
    \frac{\mu_\pi^2}{m_b^2}
    \Bigg(
        L_{i, \pi}^{(0)} + 
        L_{i, \pi}^{(1)} \frac{\alpha_s(\mu_s)}{\pi} 
    \Bigg)
    \notag \\ &
    +\frac{\mu_G^2(\mu_b)}{m_b^2}
    \Bigg(
        L_{i, G}^{(0)} + 
        L_{i, G}^{(1)} \frac{\alpha_s(\mu_s)}{\pi} 
    \Bigg)
    +\frac{\rho_D^3(\mu_b)}{m_b^3}
    \Bigg(
        L_{i, D}^{(0)} + 
        L_{i, D}^{(1)} \frac{\alpha_s(\mu_s)}{\pi} 
    \Bigg)
    \notag \\ &
    +\frac{\rho_{LS}^3(\mu_{b})}{m_b^3}
    \Bigg(
        L_{i, LS}^{(0)} + 
        L_{i, LS}^{(1)} \frac{\alpha_s(\mu_s)}{\pi} 
    \Bigg)
    +O\left(\frac{1}{m_b^4}\right)
    \Bigg]\ ,
    \label{eqn:defLi}
\end{align}
where in this case we allow for a cut on $E_l$ only.
All functions $L_i$ depend on $\rho$ and the cut $E_\mathrm{cut}$:
$L_i \to L_i(\rho,\hat E_\mathrm{cut})$.

The total semileptonic rate corresponds to  
\begin{equation}\label{eq:Gammasl}
    \Gamma_\mathrm{sl} = 
    \frac{m_b^5 G_F^2 A_\mathrm{ew}}{192 \pi^3} | V_{cb}|^2
    Q_{0,0}(\rho,0)
    =
    \Gamma_0  \,  L_0(\rho,0) ,
\end{equation}
with no cut applied, namely $E_\mathrm{cut}=q^2_\mathrm{cut} =0 $\ . For the partial decay width,
we similarly have
\begin{align}
    \Delta \Gamma_\mathrm{sl}(E_\mathrm{cut}) &=
    \Gamma_0 \, L_0(\rho, \hat E_\mathrm{cut})\ , &
    \Delta \Gamma_\mathrm{sl}(q^2_\mathrm{cut}) &=
    \Gamma_0 \, Q_{0,0}(\rho, \hat q^2_\mathrm{cut})\ .
\end{align}
The ratios defined in \eqref{eqn:defOmoments} correspond to
\begin{align}
    \langle (q^2)^n \rangle &= 
    \frac{Q_{n,0}}{Q_{0,0}}\ , &
    \langle E_l^n \rangle &=
    \frac{L_n}{L_0}\ .
    \label{eqn:momratios}
\end{align} 
The centralized moments are obtained by inserting the double expansions of \eqref{eqn:defQij} or~\eqref{eqn:defLi} into (\ref{eqn:defOmoments}-\ref{eqn:defqicentralized})
and re-expanding in $\alpha_s$ and $1/m_b$ up to the relevant order.
To assemble the $M_X^2$ moments, we express the hadronic invariant mass
in terms of the parton level quantities in the $B$ rest frame:
\begin{equation}
    M_X^2 = (M_B v -q)^2 = M_B^2 + q^2 - 2 M_B q_0\ .
\end{equation}
The moments of $M_X^2$ are obtained as linear combinations of the mixed moments $Q_{i,j}$:
\begin{align}
  M_n &=  
  \frac{1}{\Gamma_0}
  \int_{E_l \ge E_\mathrm{cut}}
  \text{d}E_l \, \text{d}q_0 \, \text{d} q^2 \,
  (M_B^2-2 M_B q_0+q^2)^n
  \frac{\text{d}^3 \Gamma}{\text{d}E_l \, \text{d}q_0 \, \text{d} q^2}  \notag \\
  &=
  \sum_{i=0}^n \sum_{j=0}^i
  \binom{n}{i}
  \binom{i}{j}
  (M_B^2)^{n-i}(-2M_B)^{i-j}
  Q_{j,i-j} \ ,
  \label{eqn:MnDefinition}
\end{align}
and $\langle (M_X^2)^n \rangle = M_n/M_0$.

In \texttt{kolya}, we first implement all building blocks introduced 
above, corresponding to the on-shell scheme for both $m_b$ and $m_c$.
We collect in the Tab.~\ref{tab:overview} the list of references 
from which the various building blocks are retrieved.
The implementation of the building blocks is described in Sec.~\ref{subsec:implementation}.
The implementation of the NNLO corrections to $E_l$ and $M_X$ moments, based on 
the results published in Ref.~\cite{Biswas:2009rb}, requires a dedicated discussion in 
sections~\ref{subsec:NNLOEl} and~\ref{subsec:NNLOMX}.
In  Sec.~\ref{subsec:tokinetic}, we perform a scheme change to the kinetic scheme to obtain 
the final prediction for the total rate and the centralized moments. 

\begin{table}
\centering
\begin{tabular}{c|ccccl}
  $\Gamma_\mathrm{sl} $ & tree  & $\alpha_s$ & $\alpha_s^2$ & $\alpha_s^3$  \\
  \cline{1-5}
  \multirow{ 2}{*}{Partonic} &
  \multirow{ 2}{*}{} & 
  \multirow{ 2}{*}{\cite{Nir:1989rm}} & 
  \multirow{ 2}{*}{\cite{Pak:2008qt,Pak:2008cp,Dowling:2008mc,Egner:2023kxw}} & 
  \multirow{ 2}{*}{\cite{Fael:2020tow}} & \\
  \multirow{ 2}{*}{$\mu_\pi^2,\mu_G^2$} &
  \multirow{ 2}{*}{\cite{Manohar:1993qn,Blok:1993va}} & 
  \multirow{ 2}{*}{\cite{Becher:2006qw,Alberti:2012dn,Alberti:2013kxa,Mannel:2015jka}} & 
  \multirow{ 2}{*}{} & & \\
  \multirow{ 2}{*}{$\rho_D^3,\rho_{LS}^3$} &
  \multirow{ 2}{*}{\cite{Gremm:1996df}} & 
  \multirow{ 2}{*}{\cite{Mannel:2021zzr}} & 
  \multirow{ 2}{*}{} & & \\
  \multirow{ 2}{*}{$1/m_b^4,1/m_b^5$} &
  \multirow{ 2}{*}{\cite{Dassinger:2006md,Mannel:2010wj,Fael:2018vsp,Mannel:2023yqf}} & 
  \multirow{ 2}{*}{} & 
  \multirow{ 2}{*}{} & & \\[10pt] 
  \cline{1-5}
  $q_n(q^2_\mathrm{cut})$ & tree  & $\alpha_s$ & $\alpha_s^2$ & 
  \\
  \cline{1-5}
  \multirow{ 2}{*}{Partonic} &
  \multirow{ 2}{*}{} & 
  \multirow{ 2}{*}{\cite{Aquila:2005hq,Mannel:2021zzr}} & 
  \multirow{ 2}{*}{\cite{Fael:2024gyw}} & 
  \multirow{ 2}{*}{} & \\
  \multirow{ 2}{*}{$\mu_G^2,\mu_\pi^2$} &
  \multirow{ 2}{*}{\cite{Manohar:1993qn,Blok:1993va}} & 
  \multirow{ 2}{*}{\cite{Alberti:2012dn,Alberti:2013kxa}} & 
  \multirow{ 2}{*}{} & & \\
  \multirow{ 2}{*}{$\rho_D^3,\rho_{LS}$} &
  \multirow{ 2}{*}{\cite{Gremm:1996df}} & 
  \multirow{ 2}{*}{\cite{Mannel:2021zzr}} & 
  \multirow{ 2}{*}{} & & \\
  \multirow{ 2}{*}{$1/m_b^4,1/m_b^5$} &
  \multirow{ 2}{*}{\cite{Fael:2018vsp,Mannel:2023yqf}} & 
  \multirow{ 2}{*}{} & 
  \multirow{ 2}{*}{} & & \\[10pt]
  \cline{1-5}
  $\ell_n(E_\mathrm{cut}), h_n(E_\mathrm{cut})$ & tree  & $\alpha_s$ & $\alpha_s^2 \beta_0$ & $\alpha_s^2$  \\
  \cline{1-5}
  \multirow{ 2}{*}{Partonic} &
  \multirow{ 2}{*}{} & 
  \multirow{ 2}{*}{\cite{Trott:2004xc,Aquila:2005hq,NNLO_El}} & 
  \multirow{ 2}{*}{\cite{Aquila:2005hq}} & 
  \multirow{ 2}{*}{\cite{Biswas:2009rb}}$*$ & \\
  \multirow{ 2}{*}{$\mu_G^2,\mu_\pi^2$} &
  \multirow{ 2}{*}{\cite{Manohar:1993qn,Blok:1993va}} & 
  \multirow{ 2}{*}{\cite{Becher:2007tk,Alberti:2013kxa}} & 
  \multirow{ 2}{*}{} & & \\
  \multirow{ 2}{*}{$\rho_D^3$} &
  \multirow{ 2}{*}{\cite{Gremm:1996df}} & 
  \multirow{ 2}{*}{} & 
  \multirow{ 2}{*}{} & & \\
  \multirow{ 2}{*}{$1/m_b^4, 1/m_b^5$} &
  \multirow{ 2}{*}{\cite{Dassinger:2006md,Mannel:2010wj,Mannel:2023yqf}} & 
  \multirow{ 2}{*}{} & 
  \multirow{ 2}{*}{} & & \\
\end{tabular} 
\caption{Schematic overview of the perturbative corrections implemented for the rate $\Gamma_\mathrm{sl}$, the $q^2$ moments, the $E_l$ and $M_X^2$ moments. (*) The $\alpha_s^2$ corrections to $h_n$ and $\ell_n$ are only available for several $\rho$ and $E_{\rm cut}$ values in \cite{Biswas:2009rb}.}
\label{tab:overview}
\end{table}

\subsection{Analytic expressions and grids for QCD corrections}
\label{subsec:implementation}

In \texttt{kolya}, the tree level expressions up to $O(1/m_b^5)$ (see 
Refs.~\cite{Fael:2018vsp,Dassinger:2006md,Mannel:2010wj,Mannel:2023yqf}) are implemented in an exact analytic form.
For example, the tree-level expressions at leading order in $1/m_b$ for $L_i^{(0)} (\rho, \hat E_\mathrm{cut})$ in \eqref{eqn:defLi} are coded in Python as follows.
\begin{minted}{python}
from numba import jit
import math

@jit(cache=True, nopython=True)
def L_0(i,elcuthat,r,dEl,dr):
    """ Tree level (partonic) for El moments and their derivatives """
    y  = 2*elcuthat
    logy = math.log((1-y)/r**2)
    # tree level function
    if (dEl == 0 and dr == 0 and i==0):
        return (1-8*r**2-6*r**4+12*logy*r**4+4*r**6-r**8
            -(2*r**6)/(-1+y)**2-(6*r**4*(1
            +r**2))/(-1+y)
            -2*r**4*(-3+r**2)*y+2*(-1+r**2)*y**3+y**4)
    if (dEl == 0 and dr == 0 and i==1):
        return (3*logy*r**4*(3+r**2)+(7-75*r**2-180*r**4
            +120*r**6-15*r**8+3*r**10)/20
            -r**6/(-1+y)**2-(r**4*(3+5*r**2))/(-1+y)
            +6*r**4*y-(r**4*(-3+r**2)*y**2)/2
            +(3*(-1+r**2)*y**4)/4+(2*y**5)/5)
    ...
\end{minted}
where the arguments of \verb|L_0| refer to the moment $i\in [0,\dots,4]$, the normalized electron energy cut $\hat E_\mathrm{cut} = E_\mathrm{cut}/m_b$ (\verb|elcuthat|)
and the mass ratio $\rho=m_c/m_b$ indicated by (\verb|r|).
The additional two arguments (\verb|dEl,dr|) are positive integers referring to the derivatives 
of $L_i(\rho, \hat E_\mathrm{cut})$ w.r.t.\ to $\hat E_\mathrm{cut}$ or $\rho$.
These derivatives are required when expressing the predictions in the kinetic scheme (see detailed discussion in Sec.~\ref{subsec:tokinetic}).
The tree level expressions up to $O(1/m_b^3)$ for the moments are implemented in the files \verb|Q2moments_SM.py|, \verb|Elmoments_SM.py| and \verb|MXmoments_SM.py|. The power corrections of order $1/m_b^4$ and $1/m_b^5$ are given in separate files \verb|Q2moments_HO.py|, \verb|Elmoments_HO.py| and \verb|MXmoments_HO.py|.
For the $q^2$ moments, the higher power corrections are in \verb|Q2moments_HO_RPI.py|

In order to use \texttt{kolya} for phenomenological applications,
for instance to perform a fit, 
it is important to ensure adequate speed for the numerical evaluations. 
To this end, our implementation utilizes the library \texttt{Numba}~\cite{Numba},
which translates Python functions to optimized machine code 
at runtime using the standard LLVM compiler library.
The functions decorated with \verb|@jit|, as shown in the example above,
are compiled to machine code ``just-in-time'' for execution at native machine-code speed. 
Numba-compiled routines in Python approach the speed of C or FORTRAN.

Let us now discuss the implementation of the QCD corrections in \texttt{kolya}. For $\Gamma_\mathrm{sl}$,
the functions in \eqref{eqn:defQij} depend only on 
the mass ratio $\rho$ and we implement the NLO corrections
up to $1/m_b^3$ using the analytic expressions given in Ref.~\cite{Mannel:2021zzr}. 
At NNLO in the free bottom quark approximation, there are
asymptotic expansions available either in the limit $\rho \to 0$~\cite{Pak:2008cp,Pak:2008qt} or $\delta = 1 -\rho \to 0$~\cite{Dowling:2008mc}. Recently, analytic expressions for the NNLO corrections written in terms of iterated integrals were presented in Ref.~\cite{Egner:2023kxw}.
In \texttt{kolya} we use the expressions expanded in terms of $\delta = 1 -\rho$ up to $\delta^{46}$ given in Ref.~\cite{Egner:2023kxw}. For the third-order correction
to $\Gamma_\mathrm{sl}$ we implement the asymptotic 
expansion up to $\delta^{12}$ computed in Ref.~\cite{Fael:2020tow}.
The total rate is implemented up to $1/m_b^3$ in \verb|TotalRate_SM.py|, the higher power corrections are given in \verb|TotalRate_HO.py| and \verb|TotalRate_HO_RPI.py|.

We use interpolation grids to implements most of the QCD corrections
for the moments.
Specifically, we use grids for all NLO corrections, the NNLO corrections to the $q^2$ moments
and the so-called BLM corrections (of order $\alpha_s^2 \beta_0$) \cite{Brodsky:1982gc} to the $E_l$ and $M_X$ moments.

Analytic expressions for the differential rate $\text{d}\Gamma/\text{d}q^2$ for a free bottom quark are
available at NLO from Ref.~\cite{Jezabek:1988iv} 
and at NNLO from Ref.~\cite{Fael:2024gyw}.
The NLO corrections to leading order in $1/m_b$ were computed also for the triple differential rate in Ref.~\cite{Aquila:2005hq,Trott:2004xc}. 
The NLO corrections to $\mu_G^2$ and $\rho_D^3$ for the 
$q^2$ spectrum have been presented in Ref.~\cite{Mannel:2021zzr}. 
By making use of reparametrization invariance~\cite{Manohar:2010sf}, one can also show that
in the on-shell scheme
\begin{align}
 -\frac{1}{2}
  Q_{i0}^{(n)}(\rho, q^2_\mathrm{cut})  &=  
  Q_{i0, \pi}^{(n)}(\rho, q^2_\mathrm{cut})\ , &
 -Q_{i0,G}^{(n)} (\rho, q^2_\mathrm{cut}) &=  
 Q_{i0, LS}^{(n)}(\rho, q^2_\mathrm{cut})\ ,
\end{align}
to all orders $n \ge 0$ in the perturbative expansion. 
Therefore, the functions $Q_{ij}$ entering \eqref{eqn:defQij}
can be computed for any $q^2_\mathrm{cut}$ and $\rho$
via:
\begin{equation}
    Q_{i0} (\rho, \hat q^2_\mathrm{cut}) = 
    \frac{1}{\Gamma_0}
    \int_{\hat q^2_\mathrm{cut}}^{(1-\rho)^2}
     (\hat q^2)^i \, \frac{\mathrm{d}\Gamma}{\mathrm{d} \hat q^2}
     \,\mathrm{d}\hat q^2\ .
     \label{eqn:integralq2}
\end{equation}
The differential rate ${\mathrm{d}\Gamma}/{\mathrm{d} \hat q^2}$ at higher orders is expressed in terms of functions defined via iterated integrals like 
harmonic polylogarithms (HPLs)~\cite{Remiddi:1999ew} and generalized polylogarithms (GPLs)~\cite{Goncharov:1998kja,Goncharov:2001iea}.
It is not convenient to integrate the differential
rate numerically ``on-the-fly'' since there are several HPLs and GPLs whose evaluation (for instance with GiNaC~\cite{Bauer:2000cp}) is time-consuming.
For this reason, we opt to implement all higher QCD corrections for the moments not in an exact form, but through Chebyshev two-dimensional grids. The functions 
implemented via grids are $Q^{(1)}_{ij}, Q^{(2)}_{ij}, Q_{ij,\pi}^{(1)}
 Q_{ij,G}^{(1)},  Q_{ij,D}^{(1)},  Q_{ij,LS}^{(1)}$ in \eqref{eqn:defQij} 
 and $L^{(1)}_{ij}, L^{(2)}_{ij}, L_{ij,\pi}^{(1)}
 L_{ij,G}^{(1)}$ in \eqref{eqn:defLi}.

Let us briefly review here how a generic function $f(x)$ can be discretized on a grid consisting of the so-called Chebyshev points (for more details see e.g.~Ref.~\cite{Press:1992zz}). The idea is to evaluate 
$f(x)$ in $n$ points $x_n$ corresponding to the zeros of the Chebyshev polynomial $T_n(x)$ of degree $n$:
\begin{equation}
    x_k = \cos \left( \frac{\pi (k-1/2)}{n} \right) \text{ for }k=1,\dots,n\ .
    \label{eqn:defxk}
\end{equation}
If $f(x)$ is an arbitrary function defined on the domain $x \in [-1,1]$, 
we calculate the coefficients $c_j$, with $j = 0, . . . , n-1$ given by
\begin{equation}
    c_j = \frac{2}{n}\sum_{k=1}^n f(x_k) T_j(x_k)\ ,
    \label{eqn:gridsCij}
\end{equation}
which can be employed to construct the polynomial
\begin{equation}
    \tilde f(x) = -\frac{1}{2}c_0 + \sum_{k=1}^{n-1} c_k T_k(x) \approx f(x)\ ,
    \label{eqn:gridsftilde}
\end{equation}
approximating $f(x)$ in the interval $[-1,1]$.
In particular $\tilde f(x) = f(x)$ for all $n$ zeros of $T_n(x)$.
In case the function to interpolate $f(y)$ is defined between two
arbitrary limits, e.g.\ $y \in [a,b]$, we apply the variable transformation
\begin{equation}
    x = \frac{y-\frac{1}{2}(b+a)}{\frac{1}{2} (b-a)}\ ,
\end{equation}
and then perform the interpolation in $x$ as before.

\begin{figure}
    \centering
    \includegraphics{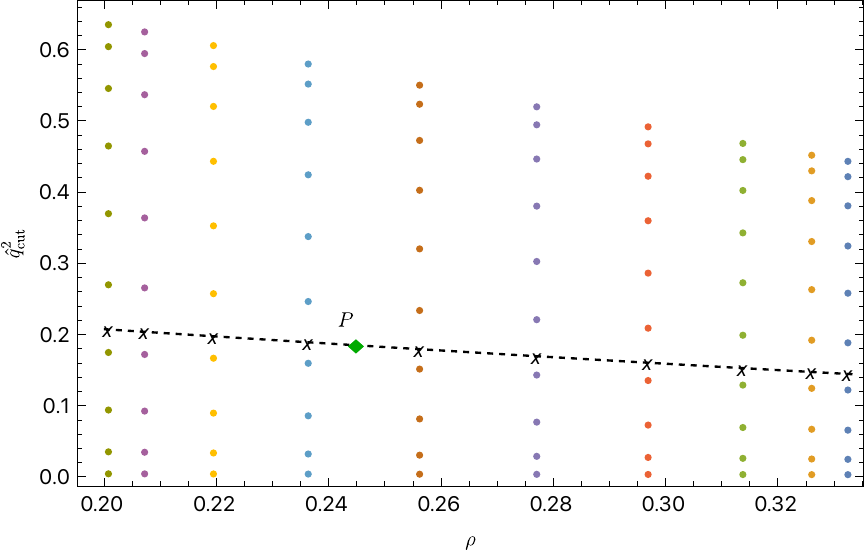}
    \caption{Example of how we discretize functions depending on two variables using Chebyshev nodes.
    In the plot, we consider the grid for $Q_{ij}$ depending on $\rho$ and $\hat q^2_\mathrm{cut}$.}
    \label{fig:example-grids}
\end{figure}

In our setup, the functions to interpolate depend on
$\rho$ and $\hat q^2_\mathrm{cut}$
(or $\rho$ and $\hat E_\mathrm{cut}$) and can be obtained via two consecutive one-dimensional Chebyshev interpolations.
First, we discretize the interval $\rho \in [1/6,1/3]$ (relevant for the phenomenology) in $n_\rho$ points distributed according to \eqref{eqn:defxk}. 
Then for each $\rho_k \in \{ \rho_1, \dots, \rho_{n_\rho} \}$, 
we discretize $\hat q^2_\mathrm{cut}$ or $\hat E_\mathrm{cut}$ into further $n_\mathrm{cut}$ points 
within the allowed range: $\hat q^2_\mathrm{cut} \in [0, (1-\rho)^2]$
or $\hat E_\mathrm{cut} \in [0, (1-\rho^2)/2]$.
An example of how the discretization is performed is shown in Fig.~\ref{fig:example-grids}, 
for a grid in $\rho$ and $\hat q^2_\mathrm{cut}$ with $n_\rho =n_{\mathrm{cut}}= 10$.

To estimate the function at a new point $P=(\rho_P, \hat q^2_P)$ 
(the green diamond in Fig.~\ref{fig:example-grids}), we proceed as follows.
For each $\rho_k,  k=1,\dots,n_\rho$, we calculate $Q(\rho_k, \hat q^2_P (1-\rho_P)^2/(1-\rho_k)^2)$
using one-dimensional interpolations in the variable $\hat q^2_\mathrm{cut}$. These values calculated
at fixed $\rho$ are shown by black crosses in Fig.~\ref{fig:example-grids}. 
Afterwards, we use them as nodes for a second interpolation this time along the $\rho$ direction, as displayed by the black dashed line in Fig.~\ref{fig:example-grids}. The second interpolation yields
the final estimate for $f(\rho_P, \hat q^2_P)$.

For the implementation of the QCD corrections
to $h_i$ and $\ell_i$, which depend
on $\rho$ and $E_\mathrm{cut}$, 
we also use Chebyshev interpolation grids.
At NLO, it is possible to write the differential rate $\text{d}\Gamma/\text{d}E_l$ in a closed
analytic form at NLO~\cite{NNLO_El} for a free quark.
To compute $Q_{ij}(\rho,\hat E_\mathrm{cut})$ and $L_i(\rho,\hat E_\mathrm{cut})$
at $O(\alpha_s)$ we perform the one variable integration,
as for instance
\begin{align}
    L_i(\rho, \hat E_\mathrm{cut}) &= 
    \frac{1}{\Gamma_0}
    \int_{\hat E_\mathrm{cut}}^{(1-\rho^2)/2}
    \, (\hat E_l)^i \,
    \frac{\mathrm{d}\Gamma}{ \mathrm{d} \hat E_l}
    \, \mathrm{d}\hat E_l\ , 
    \notag \\ 
    Q_{ij}(\rho, \hat E_\mathrm{cut}) &= 
    \frac{1}{\Gamma_0}
    \int_{\hat E_\mathrm{cut}}^{(1-\rho^2)/2} \,
    \frac{\mathrm{d} Q_{ij}}{ \mathrm{d} \hat E_l}
    \, \mathrm{d}\hat E_l\ .
\end{align}
In the last equation, we define
\begin{equation}
\frac{\mathrm{d} Q_{ij}}{ \mathrm{d} \hat E_l}
= \int 
(q^2)^i (q_0)^j
\frac{\mathrm{d}^3 \Gamma}{\mathrm{d}q^2 \, \mathrm{d}q_0  \, \mathrm{d}El}\,
\mathrm{d}q^2 \mathrm{d}q_0 
\end{equation}
which can also be computed analytically up to NLO
following Ref.~\cite{Fael:2022frj,NNLO_El}.

At order $1/m_b^2$, the NLO corrections $Q_{ij}(\rho, \hat E_\mathrm{cut})$ and $L_i(\rho, \hat E_\mathrm{cut})$
can be calculated from the triple differential distributions given in Refs.~\cite{Alberti:2012dn,Alberti:2013kxa} 
by performing the phase space integration numerically as described in Ref.~\cite{Alberti:2016fba}.
The NLO corrections to $\rho_D^3$ and $\rho_{LS}^3$ are not known at the moment. 

The values of the coefficients $c_j$ in \eqref{eqn:gridsCij} for all grids are stored in the directory \verb|grids| as multidimensional 
arrays. The routines which perform the interpolation of the NLO and NNLO corrections at $O(1/m_b^0)$ are implemented in \verb|NLOpartonic.py| and \verb|NNLOpartonic.py|. The routines for the NLO corrections to the power-suppressed terms are given in \verb|NLOpw.py|.

We validate the grid implementation by generating 100 random points in the two-dimensional plane $(\rho,\hat q^2_\mathrm{cut})$ or $(\rho,\hat E_\mathrm{cut})$.
For each point, we compare the approximation provided by the grids and high-precision evaluations
obtained with Mathematica. 
We verify that the two estimates differ by less than $10^{-5}$ for the points considered.

\subsection{NNLO corrections to the lepton energy moments}
\label{subsec:NNLOEl}
The NNLO corrections to the $\ell_i$ moments are not known in a closed form. As discussed, the BLM corrections are implemented through interpolation grids. The remaining ``non-BLM'' corrections are only known for specific values of $\rho$ and $E_{\rm{cut}}$ from Ref.~\cite{Biswas:2009rb}. Their functional form can be obtained from a two-dimensional fit to these data points. In order to perform this fit, we write:
\begin{align}
    \ell_n(\rho, E_{\rm{cut}})=&\ (m_b)^n\Bigg[Y_n^{(0)}+Y_n^{(1)}\frac{\alpha_s(\mu_s)}{\pi}+\Big(\beta_0Y_n^{(2,\rm{BLM})}+Y_n^{(2,\rm{nonBLM})}\Big)\left(\frac{\alpha_s(\mu_s)}{\pi}\right)^2\nonumber\\
    &+O\left(\frac{1}{m_b},\left(\frac{\alpha_s(\mu_s)}{\pi}\right)^3\right)\Bigg]\ ,
\end{align}
where the $E_{\rm{cut}}$- and $\rho$-dependence of $Y_n^{(i)}$ is implied and $Y_n^{(0)}$ is the partonic contribution without any $\alpha_s$ or $1/m_b$ corrections. In terms of the building-blocks defined in \eqref{eqn:defLi}, we can write the non-BLM terms as:
\begin{align}
    Y_1^{(2,\rm{nonBLM})}&=\frac{L_1^{(2,\rm{nonBLM})}}{L_0^{(0)}}-\frac{L_0^{(2,\rm{nonBLM})}L_1^{(0)}}{(L_0^{(0)})^2}\ ,\nonumber\\
    Y_2^{(2,\rm{nonBLM})}&=\frac{L_2^{(2,\rm{nonBLM})}}{L_0^{(0)}}-2\frac{L_1^{(2,\rm{nonBLM})}L_1^{(0)}}{(L_0^{(0)})^2}+L_0^{(2,\rm{nonBLM})}\left(2\frac{(L_1^{(0)})^2}{(L_0^{(0)})^3}-\frac{L_2^{(0)}}{(L_0^{(0)})^2}\right)\ ,\nonumber\\
    Y_3^{(2,\rm{nonBLM})}&=\frac{L_3^{(2,\rm{nonBLM})}}{L_0^{(0)}}-3\frac{L_2^{(2,\rm{nonBLM})}L_1^{(0)}}{(L_0^{(0)})^2}+3L_1^{(2,\rm{nonBLM})}\left(2\frac{(L_1^{(0)})^2}{(L_0^{(0)})^3}-\frac{L_2^{(0)}}{(L_0^{(0)})^2}\right)\nonumber\\
    &\hspace{0.5cm}+L_0^{(2,\rm{nonBLM})}\left(-6\frac{(L_1^{(0)})^3}{(L_0^{(0)})^4}+6\frac{L_1^{(0)}L_2^{(0)}}{(L_0^{(0)})^2}-\frac{L_3^{(0)}}{(L_0^{(0)})^2}\right)\ .
\end{align}
Ref.~\cite{Biswas:2009rb} gives the $L_n^{(2,\rm{nonBLM})}$ terms at $\rho = \{0.20, 0.22, 0.24, 0.25, 0.26, 0.28\}$ and $y \equiv 2\hat{E}_{\rm{cut}} =\{0, 0.1, \ldots, 0.7\}$. From these\footnote{Note that the $L_i^{(n)}$ defined  in Ref.~\cite{Biswas:2009rb} are normalized to the total partonic rate without cut while we only normalize to $\Gamma_0$ defined in \eqref{eq:Gammasl}.}, the non-BLM contributions $Y_n^{(2,\rm{nonBLM})}$ to the $\ell_n$ moments are obtained by combining with the tree-level building blocks $L_n^{(0)}$. In Ref.~\cite{Gambino:2011cq}, these non-BLM contributions are studied in detail and compared to the effect of their BLM counterparts. 
We fit the values for $Y_n^{(2,\rm{nonBLM})}$ assuming the 
following polynomial ansatz
\begin{equation}
    Y_n(\rho,y)=
    \sum\limits_{i=1}^5(a_{n,i}+b_{n,i}\rho)(y+\rho^2-1)^i\ ,\label{eq:fitform}
\end{equation} 
for each moment $n$. Following  Ref.~\cite{Gambino:2011cq}, we only include one power of $\rho$ in our ansatz, but keep up terms up $y^5$ in our interpolating fit. In addition, the ansatz is chosen to ensure that the non-BLM corrections vanish at the end point $y=1-\rho^2$ \cite{Biswas:2009rb}. We stress that our approach differs from \cite{Gambino:2011cq} as we first construct $Y_n^{(2,\rm nonBLM)}$ in each available $(\rho, y)$ point and then perform the analysis. Fitting first the $L_n^{(2,\rm{nonBLM})}$ and then combining them resulted in strongly oscillating functions due to accidental cancellations. Fitting directly $Y_n$, we find
\begin{align}
    Y_1&^{(2,\rm{nonBLM})}(\rho,y)=\ 
    \notag\\ &
    (72.57 \rho -25.62) \left(y +\rho ^2-1\right)^5+(177.60 \rho -64.65) \left(y +\rho ^2-1\right)^4\nonumber\\
    &+(157.19 \rho -59.27) \left(y
   +\rho ^2-1\right)^3+(62.69 \rho -24.32) \left(y +\rho ^2-1\right)^2\nonumber\\
   &+(11.25 \rho -4.35) \left(y +\rho ^2-1\right)\nonumber\ ,\\
   Y_2&^{(2,\rm{nonBLM})}(\rho,y)=\ 
   \notag \\ &
   (12.61\, -44.47 \rho ) \left(y +\rho ^2-1\right)^5+(32.28\, -112.34 \rho ) \left(y +\rho ^2-1\right)^4\nonumber\\
   &+(29.36\, -100.36 \rho )
   \left(y +\rho ^2-1\right)^3+(11.07\, -36.89 \rho ) \left(y +\rho ^2-1\right)^2\nonumber\\
   &+(1.42\, -4.55 \rho ) \left(y +\rho ^2-1\right) \ ,
   \end{align}
and
   \begin{align}
   Y_3&^{(2,\rm{nonBLM})}(\rho,y)=\ 
   \notag \\ &
   (32.23 \rho -7.28) \left(y +\rho ^2-1\right)^5+(79.34 \rho -17.72) \left(y +\rho ^2-1\right)^4\nonumber\\
   &+(67.06 \rho -14.67) \left(y
   +\rho ^2-1\right)^3+(21.52 \rho -4.49) \left(y +\rho ^2-1\right)^2\nonumber\\
   &+(1.71 \rho -0.289) \left(y +\rho ^2-1\right)\ .
   \label{eqn:nonBLMEl}
\end{align} 
These functions are implemented in \texttt{kolya}. We do not assign any uncertainty to these fit functions. 

\begin{figure}
    \centering
    \includegraphics[width=0.48\linewidth]{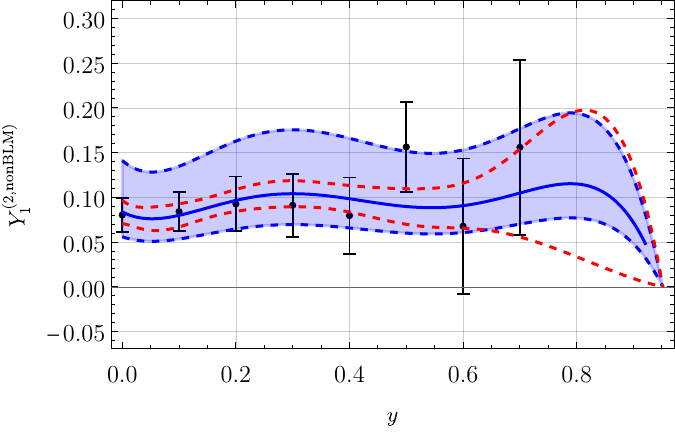}
    \includegraphics[width=0.48\linewidth]{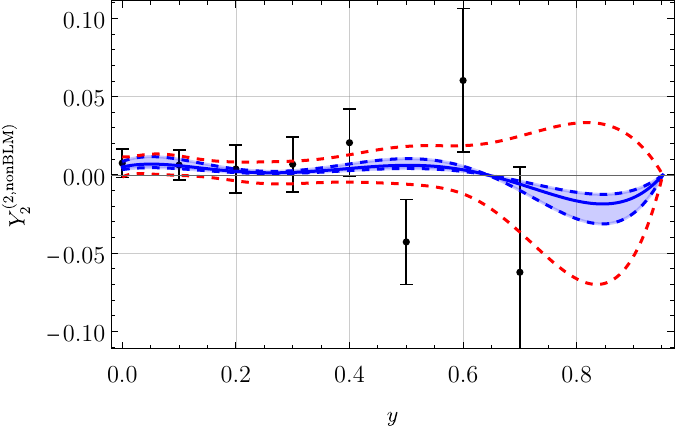}
    \includegraphics[width=0.48\linewidth]{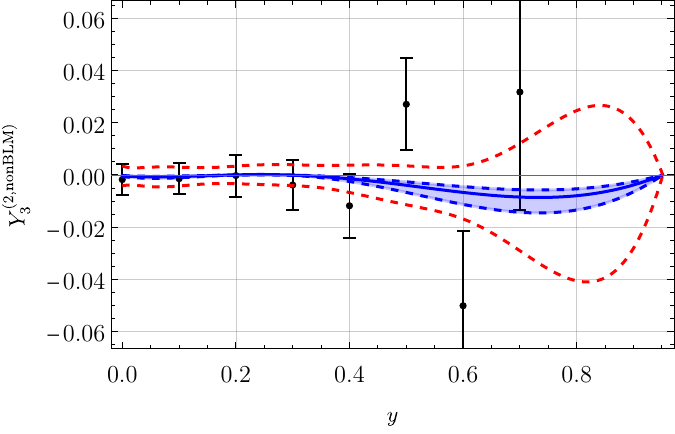}
    \caption{
    Results of our fitted non-BLM contributions $Y_n^{(2,\rm{nonBLM})}$ (solid blue line) as a function of $y=2E_\mathrm{cut}/m_b$ for $\rho=0.22$ as discussed in the text. Data points are constructed from \cite{Biswas:2009rb}. 
    }
    \label{fig:nonBLM}
\end{figure}

In Fig.\ \ref{fig:nonBLM}, we show our fit results for $Y_{1,2,3}^{(2,\rm{nonBLM})}$, 
as a function of $y$ for fixed $\rho = 0.22$ (solid blue line). Due to the fit ansatz, we observe a light oscillatory behavior as a function of $y$. In black, we show the constructed data points at fixed $y$ obtained from \cite{Biswas:2009rb}. The fit uncertainty is given by the red dotted line, which represents the $90\%$ C.L.\ interval of the fit. Since we only have data points up to $y=0.7$ and impose that the contribution vanishes at the endpoint, we notice large uncertainties towards higher $y$ values. In a typical phenomenological analysis, missing higher order terms (here $\alpha_s^3$) terms would be accounted for by varying the scale of $\alpha_s$. The blue bland corresponds to the effect of such a scale variation from $[\alpha_s(m_b^{\rm kin}/2), \alpha_s(2m_b^{\rm kin})]$. For $Y_1$, we observe that the $\alpha_s$ variation covers the data points and fit uncertainty. For the higher moments, we observe that the fit uncertainty is higher than the $\alpha_s$ variation for large $y$. However, given the smallness of the BLM contributions to these moments, our default setting is to not include an additional uncertainty for these corrections.

\subsection{\boldmath NNLO corrections to the hadronic invariant mass moments \unboldmath}
\label{subsec:NNLOMX}
The nonBLM corrections to the hadronic moments can be obtained from Ref.~~\cite{Aquila:2005hq}. However, as for the lepton energy moments, the additional ``non-BLM'' contributions are only known numerically for several values of $\rho$ and $E_{\rm cut}$~\cite{Biswas:2009rb}. In Ref.~\cite{Biswas:2009rb}, building blocks $H_{ij}$ are defined as
\begin{align}
    H_{ij}(\rho,\hat E_{\rm cut})=\frac{1}{\Gamma_0L_0^{(0)}(\rho,0)}\int_{ E_{\rm cut}}\text{d}\hat E_l\, \text{d}\hat q_0\,\text{d}\hat q^2\, \left(\frac{m_x^2-m_c^2}{m_b^2}\right)^i\left(\frac{E_h}{m_b}\right)^j \frac{\text{d}^3\Gamma}{\text{d}\hat E_l\, \text{d}\hat q_0\,\text{d}\hat q^2}\ ,
\end{align}
where $L_0^{(0)}(\rho,0)$ is defined in \eqref{eqn:defLi} and where $m_x$ and $E_x$ are partonic invariant mass and energy. A linear combination of these $H_{ij}$ corresponds to our building blocks $Q_{ij}$ defined in \eqref{eqn:MnDefinition}. This then allows us to calculate the non-BLM contributions to $h_1$ by combining different $H_{ij}$ contributions. However, not all the $Q_{ij}$ moments necessary to construct the hadronic moments $h_2$ and $h_3$ are calculated. In terms of the definition of \cite{Biswas:2009rb}, the non-BLM contributions of the combinations
\begin{align}
    H_{20},\ H_{30},\ H_{21}\ ,\nonumber
\end{align}
are missing. 

\begin{figure}
    \centering
    \includegraphics[width=0.48\linewidth]{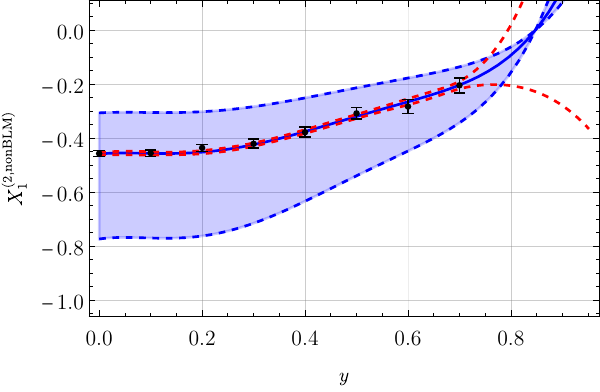}
    \includegraphics[width=0.48\linewidth]{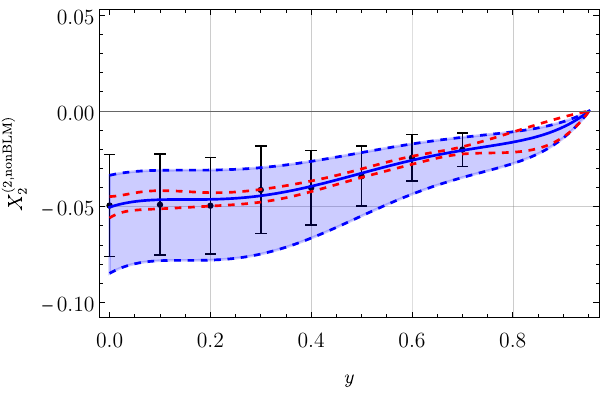}
    \includegraphics[width=0.48\linewidth]{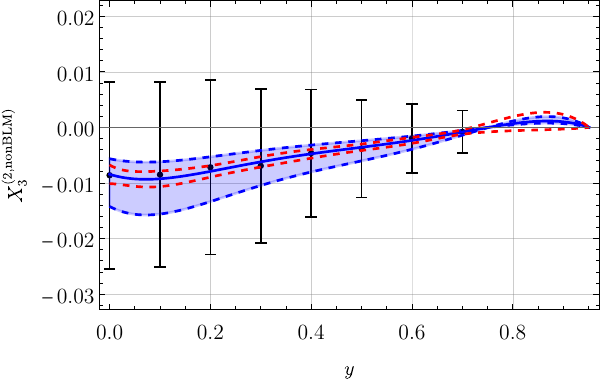}
    \caption{The non-BLM contributions $X_n^{(2,\rm{nonBLM})}$ as a function of $y=2E_\mathrm{cut}/m_b$ for $\rho=0.22$ as described in the text.}
    \label{fig:MXnonBLM}
\end{figure}

In order to determine the effect of the non-BLM terms, we write 
\begin{align}
    h_n(y,\rho) = X_n(y,\rho)+m_b^{2}\left(\frac{\alpha_s(\mu_s)}{\pi}\right)^2 \left[\beta_0X_n^{(2,\rm BLM)}(y,\rho)+X_n^{(2,\rm nonBLM)}(y,\rho)\right]\ ,
\end{align}
where $X_n^{(2,\rm nonBLM)}$ is the non-BLM contribution, $X_n^{(2,\rm BLM)}$ the BLM contribution and $X_n$ contains all other contributions. For $h_1$, we can now proceed as for the lepton energy moments. Here, we use 
\begin{align}\label{eq:hnans}
    X_1^{(2,\rm nonBLM)}(\rho,y)=
    \sum\limits_{i=0}^5(a_{1,i}+b_{1,i}\rho)y^i\ ,
\end{align}
as our fit ansatz, where we do not require that the first moment vanishes at the end point. Using the data points from Ref.~\cite{Biswas:2009rb} (to be conservative, we assume a $1\%$ uncertainty on all data points for which no uncertainty was given), we then find
\begin{align}\label{eq:chi_1}
    X_1^{(2,\rm nonBLM)}(y,\rho)=&\ y ^5 (33.13\, -97.49 \rho )+y ^4 (224.82\rho -71.16)\nonumber\\
    &+y ^3 (49.01\, -160.90 \rho )+y ^2 (37.72 \rho -10.92)\nonumber\\
    &+y  (0.73\, -2.56 \rho )+2.08 \rho -0.92\ .
\end{align}
In Fig.\ \ref{fig:MXnonBLM}, we show $X_1^{(2,\rm nonBLM)}$ as a function of $y$. As for the lepton energy moments, the red dotted line shows the fit uncertainty, which now diverges close to the end point. However, as usually the experimental data does not have lepton energy cuts $y>0.8$ this feature does not pose a great problem. In addition, we notice the small uncertainty on the data points, clearly much smaller than the effect of the $\alpha_s$ variation (shown by the blue band). As such, we do not implement an additional fit uncertainty on \eqref{eq:chi_1} and implement this function into \texttt{kolya}. A similar fit was done in \cite{Gambino:2010jz}, where also $\rho^2$ terms were taken into account. We do not notice an improvement in the fit when including such terms and therefore use our minimal fit ansatz presented above. 

For the $h_2$ and $h_3$ moments, we currently do not have sufficient information to perform a fit. However, for $y=0$, we can use the expression for $Q_{ij}$ from \cite{Fael:2022frj}, where analytic expression in terms of $\delta = 1-\rho$ were given. From these, we can then also determine the non-BLM effect over the BLM contribution. We find stable results for the different $X_n$. Using $\rho=0.25$ for illustration, we have

\begin{align}
    \frac{X_1^{(2,\rm{nonBLM})}}{\beta_0X_1^{(2,\rm{BLM})}}=-0.237\ , \quad
    \frac{X_2^{(2,\rm{nonBLM})}}{\beta_0X_2^{(2,\rm{BLM})}}=-0.227\ , \quad
    \frac{X_3^{(2,\rm{nonBLM})}}{\beta_0X_3^{(2,\rm{BLM})}}=-0.241\ .
\end{align}
We do not quote an uncertainty on these values, which could be obtained from \cite{Fael:2022frj} by consider the effect of the highest power in the $\delta$ expansion. At the moment, we do not have more information for the $X_2$ and $X_3$ contributions at nonzero values of the lepton cut $y$. In \cite{Gambino:2011cq}, the non-BLM/BLM ratios of the available relevant combinations of $H_{ij}$ moments were calculated. For the values in \cite{Biswas:2009rb}, it was found that this ratio is rather independent of $y$. This may indicate that also the missing terms are constant when considering ratios of non-BLM versus BLM corrections. In \cite{Gambino:2011cq}, this constant behavior was therefore assumed for the missing $H^{(2)}_{20,21,30}$ moments\footnote{Using \cite{Fael:2022frj}, we can calculate for the first time also the ratio of $H_{20}^{(2,\rm non-BLM)}/H_{20}^{(2,BLM)}$ and similarly for $\{21\}$ and $\{30\}$. We confirm within uncertainties the ratios quoted in \cite{Gambino:2011cq}.} . On the other hand, we note that the analytic results in \cite{Fael:2024gyw} for the $q^2$ moments with a $q^2$ cut do not show a constant non-BLM over BLM behavior. 

Nevertheless, to further study $h_2$ and $h_3$, we follow \cite{Gambino:2011cq} and assume constant non-BLM over BLM ratios for the missing moments. We then find the constructed data points for  $X_2^{(2,\rm nonBLM)}$ and $X_3^{(2,\rm nonBLM)}$ given in Fig.~\ref{fig:MXnonBLM}. In blue, we present the fit to the data using a similar ansatz as in \eqref{eq:fitform} (ensuring also that the contribution vanishes at the end point). We note that the fit uncertainty (dotted red line) is very small and that also the $\alpha_s$ variation (blue band) is much smaller than the uncertainty on the data points. In addition, the assumption for the missing moments may bias our predictions. Therefore, we decide for the moment not to include any non-BLM corrections to the $h_2$ and $h_3$ hadronic mass moments.

\subsection{\boldmath Kinetic and $\overline{\mathrm{MS}}$ scheme \unboldmath}
\label{subsec:tokinetic}

In the HQE, the expressions for the observables
are expressed in terms of a short-distance mass scheme
for the bottom and charm quark. This ensures the cancellation of the leading renormalon divergence 
which arises if the observables are written in terms of the pole mass for the heavy quark~\cite{Beneke:1994sw,Bigi:1994em}.
In \texttt{kolya}, we use the so-called kinetic scheme~\cite{Bigi:1996si} for the bottom quark 
and the HQE operators.
In this scheme, the bottom quark pole mass 
is rewritten in terms of the kinetic mass using the relation
\begin{equation}
    m_b^\mathrm{pole} = 
    m_b^\mathrm{kin}(\mu) + [\overline \Lambda (\mu)]_\mathrm{pert}
    +\frac{[\mu_\pi^2(\mu)]_\mathrm{pert}}{2 m_b^\mathrm{kin}(\mu)}
    +O\left( \frac{1}{m_b^2} \right),
    \label{eqn:mpole2mkin}
\end{equation}
where the scale $\Lambda_\mathrm{QCD} \ll \mu \ll m_b$ is a Wilsonian cutoff scale, usually taken to be 1 GeV.
The latter two terms in \eqref{eqn:mpole2mkin} denoted with ``pert'' are the perturbative versions of the HQE parameters
determined from the Small Velocity sum rules~\cite{Bigi:1994ga}. 
Their explicit expressions up to order $\alpha_s^3$ are given in Refs.~\cite{Czarnecki:1997sz,Fael:2020njb}. We stress that in our implementation, we follow \cite{Fael:2020njb} and include also the decoupling effects of the charm quark in the kinetic scheme. We show the effect of this decoupling numerically in Sec.~\ref{sec:numres}. At the same time, the HQE parameters entering the expansion in ~\eqref{eqn:defQij} and~\eqref{eqn:defLi} must also be redefined in the kinetic scheme by subtracting from their perturbative contribution:
\begin{align}\label{eq:rhodpert}
    \mu_\pi^2(0) &= \mu_\pi^2(\mu) - [\mu_\pi^2(\mu)]_\mathrm{pert}\ , &
    \rho_D^3(0) &= \rho_D^3(\mu) - [\rho_D^3(\mu)]_\mathrm{pert}\ .
\end{align}
These expressions actually refer to the value of $\mu_\pi^2$ and $\rho_D^3$ in the infinite $m_b$ limit
while the fits~\cite{Bordone:2021oof,Bernlochner:2022ucr} employ definitions of the HQE parameters at finite $m_b$.
In both cases, the setup neglects (unknown) terms of $O(\mu^3)$ in $[\mu_\pi^2(\mu)]_\mathrm{pert}$ and $[\mu_G^2(\mu)]_\mathrm{pert}$.
Since the operator bases in Refs.~\cite{Bordone:2021oof} and \cite{Bernlochner:2022ucr} differ
in particular for the definition of $\mu_G^2$, a mismatch of order $\alpha_s \times \mu^3$ appears when comparing
the two frameworks. Both bases are implemented in \texttt{kolya}. 

Finally, for the charm mass, we use the $\overline{\mathrm{MS}}$
scheme. This is related to the pole charm mass via
\begin{multline}
m_c^\mathrm{pole} = 
\overline{m}_c(\mu_c) 
\left[ 
1 
+ \frac{\alpha_s(\mu_c)}{\pi}
\left(\frac{4}{3} + L \right) \right. \\
\left.+ \left(\frac{\alpha_s(\mu_c)}{\pi}\right)^2
\left( 
\frac{37}{24}L^2
+\frac{415}{72} L
+\frac{779}{96}
+\frac{\pi ^2}{6}
-\frac{\zeta_3}{6}
+\frac{1}{9} \pi ^2 \log (2)
\right)
\right]
+O(\alpha_s^3),
\label{eqn:mpole2mMS}
\end{multline}
where $L = \log(\mu_c^2/\overline{m}_c^2(\mu_c))$.

In order to implement the scheme change for the quark masses
and the HQE parameters, we replace $m_b^\mathrm{pole}$ and $m_c^\mathrm{pole}$ in the expressions for the centralized moments using \eqref{eqn:mpole2mkin} and~\eqref{eqn:mpole2mMS} and expand the formulas as a series in $\alpha_s$ and $1/m_b$.

Hard coding these scheme changes would lead to huge expressions for the moments. To illustrate this,
we write the mass conversion formulas schematically
as follows:
\begin{align}
   m_c^\mathrm{pole} &=
   \overline{m}_c (1 + \alpha_s \delta_{m_c,1} 
   +\alpha_s^2 \delta_{m_c,2} 
   +\alpha_s^3 \delta_{m_c,3} )\ ,
    \notag \\ 
    m_b^\mathrm{pole} &=
    m_b^\mathrm{kin} (1
    + \alpha_s \delta_{m_b,1}
    + \alpha_s^2 \delta_{m_b,2}
    + \alpha_s^3 \delta_{m_b,3}
    )\ .
\end{align}
The coefficients appearing at order $\alpha_s^n$ from the bottom and charm mass scheme conversion formulas are denoted by $\delta_{m_b,n}$ and $\delta_{m_c,n}$, respectively. 
Let us now consider a simple example with a function $f(\rho)$
depending only on the mass ratio. The scheme change for $f(\rho)$ up to $O(\alpha_s)$ reads
\begin{align}
    f(\rho) &= f\left(\frac{m_c^\mathrm{pole}}{m_b^\mathrm{pole}}\right)
    =f\Big(\overline{m}_c (1+ \alpha_s \delta_{m_c,1} )
    \Big/
    m_b^\mathrm{kin} (1+ \alpha_s \delta_{m_b,1})
    \Big) \notag \\ &
    = f\left(\frac{\overline{m}_c}{m_b^\mathrm{kin}}\right) 
    +\alpha_s 
    f'\left(\frac{\overline{m}_c}{m_b^\mathrm{kin}}\right) 
    \frac{\overline{m}_c}{m_b^\mathrm{kin}}
    (\delta_{m_c,1}-\delta_{m_b,1})
    +O(\alpha_s^2)\ .
    \label{eqn:masschangeexample}
\end{align}
The function $f(\rho)$ and its derivative are the building blocks entering in \texttt{kolya}. This has several advantages with respect to hard coding the expression in.

In fact for the centralized moments, the role of $f(\rho)$ is played by the building blocks denoted by $Q_{ij}$ and $L_i$ on the r.h.s.\ of \eqref{eqn:defQij} and~\eqref{eqn:defLi}.
They appear multiple times after the re-expansion of the 
ratios in \eqref{eqn:defOmoments} and the scheme change. 
Therefore, we find more convenient to calculate first all
building blocks in~\eqref{eqn:defQij} and~\eqref{eqn:defLi}
 (and the necessary derivatives) and cache 
 the results at a given value of $\rho$ and the cut.
Afterwards, we assemble the centralized moments using expressions similar to \eqref{eqn:masschangeexample}. 
This approach yields code with a much smaller size  
and improved evaluation time.

The coefficients $\delta_{m_b,n}$ and $\delta_{m_c,n}$
which enter in the scheme change are implemented
up to $O(\alpha_s^3)$ in the file \verb|schemechange_KINMS.py|. This allows in principle
to easily adopt a different mass scheme from our
default one, by changing the expression for $\delta_{m_b,n}$ and $\delta_{m_c,n}$ to the required mass scheme in a separate file.

\subsection{Numerical results for lepton energy and hadronic mass moments}\label{sec:numres}
 The accuracy of our moment predictions is summarized in Tab.~\ref{tab:overview}. As the NNLO corrections to the $q^2$ moments are known analytically \cite{Fael:2024gyw}, and a detail discussion was given recently in that reference, we do not discuss these here in detail. However, since our implementation for the lepton energy and hadronic mass moments depends on our fit for the non-BLM contributions, it is interesting to give the relative contributions to the moments. These contributions are given in Tab.~\ref{tab:contris}, where we use $E_{\rm cut}=1$ GeV. For the HQE parameters, we use $\mu_\pi^2 = 0.4\ {\rm GeV}^2$ , $\mu_G^2 = 0.35\ {\rm GeV}^2$ , $\rho_{LS}^3 = -0.15\ {\rm GeV}^3$ , $\rho_D^3 = 0.2\ {\rm GeV}^3$. For the other input parameters, we employ $m_b^\mathrm{kin}(1\mathrm{\ GeV})=4.563\mathrm{\ GeV},\ \overline{m}_c(3\mathrm{\ GeV})=0.989\mathrm{\ GeV},\ \alpha_s(m_b^\mathrm{kin})=0.2182$. 

In Tab.~\ref{tab:contris}, we also explicitly give the effect of including the charm decoupling in the kinetic scheme conversion. This contribution is labelled as $\Delta_c$. Our results are in good agreement with Ref.~\cite{Gambino:2011cq} up to the non-BLM corrections. The $\Delta_c$ effects were not included in that reference. Currently, these contributions are automatically included at NNLO in \texttt{kolya}, and they cannot be separated from other contributions. 

We observe that $\Delta_c$ has a $0.1\%$ effect and is of similar in size but opposite in sign as the non-BLM correction. Overall, the $\alpha_s^2$ contributions have a $0.05\%$ effect. 

We provide additional numerical examples and comparisons with literature in the GitLab repository, in the Jupyter notebook \texttt{example-reproduce\_literature.ipynb}.
\begin{table}
    \centering
     \begin{tabular}{c|c|c|c|c|c|c}
      Moment & tree & $\alpha_s$ & $\alpha_s^2 \beta_0$ & $\alpha_s^2$ scheme &$\alpha_s^2$ non-BLM &$\alpha_s^2$ $\Delta_c$ \\
        \hline\hline
        $\ell_1$ [GeV]  & 1.5650  & 1.5521 & 1.5540 & 1.5459  & 1.5480  &1.5465   \\
    $\ell_2$ [GeV$^2$] & 0.0895  & 0.0870  & 0.0881  & 0.0861  & 0.0867   & 0.0863 \\
    $\ell_3$ [GeV$^3$] & -0.0018  & -0.0003  &0.0004  & 0.0006  & -0.0006  & -0.0006   \\
    \hline
       $h_1$ [GeV$^2$]  & 4.166   & 4.331  & 4.304   & 4.417  & 4.381  &4.403  \\ 
    $h_2$ [GeV$^4$] & 0.609  & 0.818  & 1.001  & 0.987  & -& 0.990 \\
    $h_3$ [GeV$^6$] & 5.071  & 4.810  & 4.487  & 4.641  & -& 4.640   \\
    \end{tabular}
    
    \caption{Contributions of the different orders to the moments at $E_{\rm cut}=1$ GeV for $\mu_\pi^2 = 0.4\ {\rm GeV}^2$ , $\mu_G^2 = 0.35\ {\rm GeV}^2$ , $\rho_{LS}^3 = -0.15\ {\rm GeV}^3$ , $\rho_D^3 = 0.2\ {\rm GeV}^3$. For the other input parameters, we employ
    $m_b^\mathrm{kin}(1\mathrm{\ GeV})=4.563\mathrm{\ GeV},\ \overline{m}_c(3\mathrm{\ GeV})=0.989\mathrm{\ GeV},\ \alpha_s(m_b^\mathrm{kin})=0.2182$. }
    \label{tab:contris}
\end{table}

\section{Extension to physics beyond the SM}
\label{sec:NP}
In \texttt{kolya}, we also implement NP effects in $b\to cl\bar{\nu}_l$ decays following \cite{Fael:2022wfc}. In order to parametrize effects beyond the SM, we use 
the weak effective theory (WET), an effective
field theory valid below the EW scale written as an
expansion in powers of the inverse electroweak scale 
$G_F = 1/(\sqrt{2}v^2)$ .
The effective Hamiltonian relevant for 
$B \to X_c l \bar \nu_l$ is given by
\begin{align}
    \mathcal{H}_{\text{eff}} &= \frac{4 G_F V_{cb}}{\sqrt{2}} 
    \left [ \left(1+ C_{V_L} \right) \mathcal{O}_{V_L} 
    + \sum_{i = V_R, S_L, S_R, T} C_{i} \, \mathcal{O}_{i}  \right],
    \label{eq::Hamiltonian-NP}
\end{align}
where we consider only the dimension-six operators which contribute to the differential rate at tree level:
\begin{align}
    \mathcal{O}_{V_{L(R)}} &= \left( \bar{c} \gamma_\mu P_{L(R)} b \right) \left(\bar{\ell} \gamma^\mu P_{L} \nu_\ell \right) \, , \notag  \\
    \mathcal{O}_{S _{L(R)}} &= \left(\bar{c} P_{L(R)} b \right) \left(\bar{\ell} P_{L} \nu_\ell \right) \, , \notag \\
    \mathcal{O}_{T} &= \left(\bar{c} \, \sigma_{\mu \nu} P_{L} b \right) \left(\bar{\ell} \, \sigma^{\mu \nu} P_{L} \nu_\ell \right) \, .
\end{align}
We define $P_{L(R)} = 1/2 \, (1 \mp \gamma_5)$ and $\sigma^{\mu \nu} = \frac{i}{2} [\gamma^\mu, \gamma^\nu]$. In the SM, only $O_{V_{L}}$ contributes. We have written out this contribution explicitly, such that all Wilson coefficients $C_i$ are zero in the SM. We assume that all
$C_i$ are real-valued. We do not consider interactions with right-handed neutrinos. 

If one would like to parametrize NP effects in the SMEFT framework~\cite{Grzadkowski:2010es}, 
there would be an additional expansion in powers of $1/\Lambda$, where
$\Lambda$ corresponds to the NP scale above the EW scale. 
The tree-level matching of SMEFT operators onto the effective Hamiltonian
can be obtained from Ref.~\cite{Aebischer:2015fzz}.
In the WET, the expansion parameter is $1/v$, therefore from the SMEFT point of view
the Wilson coefficients in \eqref{eq::Hamiltonian-NP} would be further suppressed by the small ratio $(v/\Lambda)^2$.

The NP contributions to the differential rate 
of $B \to X_c l \bar \nu_l$ have been presented
in Ref.~\cite{Fael:2022wfc}.
The NP effects for the moments defined in \eqref{eqn:defOmoments}, obtained from the integration of the triple differential rate, 
can be written schematically in the following way
\begin{align}
 \langle O \rangle &= \xi_{\text{SM}} +  |C_{V_{R}}|^2 \, \xi^{\braket{V_R,V_R}}_{\text{NP}} +  |C_{S_{L}}|^2 \, \xi^{\braket{S_L,S_L}}_{\text{NP}} +|C_{S_{R}}|^2 \, \xi^{\braket{S_R,S_R}}_{\text{NP}} + |C_{T}|^2 \, \xi^{\braket{T,T}}_{\text{NP}}  \nonumber \\ 
 &  
 + \text{Re}(( C_{V_{L}} -1) C_{V_R}^*) \, \xi^{\braket{V_L,V_R}}_{\text{NP}}  
 + \text{Re}(C_{S_{L}} C_{S_R}^*) \, \xi^{\braket{S_L , S_R}}_{\text{NP}} 
 + \text{Re}(C_{S_{L}} C_{T}^*) \, \xi^{\braket{S_L , T}}_{\text{NP}} \nonumber \\
 & 
 + \text{Re}(C_{S_{R}} C_{T}^*) \, \xi^{\braket{S_R , T}}_{\text{NP}} \ .
 \label{eq::moments}
\end{align}
The coefficients denoted by $\xi$ correspond to the 
various interference terms between different effective operators. They depend on the bottom and charm quark masses, the HQE parameters and the lepton energy cut or the $q^2$ cut. 
Ref.~\cite{Fael:2022wfc} provides results for the 
power corrections at tree level up to $O(1/m_b^3)$
and NLO perturbative corrections to leading order 
in the power expansion ($1/m_b^0$).
In \eqref{eq::moments}, we assume that the NP Wilson coefficients $C_i$ are much smaller than unity therefore, when calculating the moments in \eqref{eqn:defOmoments} we expand up to quadratic NP couplings. We note that the contribution $C_{V_L} \xi^{\braket{V_L}}_{\text{NP}}$ drops out for such normalized moments while for the branching ratio, it leads to a rescaling of $V_{cb}$.

The first term $\xi_{\rm SM}$ in \eqref{eq::moments} 
corresponds to the SM prediction, whose implementation
has been described in the previous sections. The additional NP contributions generated by the effective 
operators have been implemented in \texttt{kolya}
including power corrections up to $1/m_b^3$ and NLO 
perturbative QCD corrections at partonic level.
The implementation closely follows the methods described
for the SM case. Namely, we implement the tree-level 
contributions in an exact form, while for the QCD corrections, we generate interpolation grids for their 
fast evaluation.
The NP contributions to the moments are implemented in \verb|Q2moments_NP.py|, \verb|Elmoments_NP.py| and \verb|MXmoments_NP.py|.
The NP extension of the total rate is in \verb|TotalRate_NP.py|.

\section{Usage of the library}
\label{sec:usage}

\subsection{Installation } 
The software \texttt{kolya} requires Python version 3.6 or above and runs on Linux and Mac.
The code is released under the GNU GPL v3 license.
To download the package, clone the master branch of the Git repository via
\begin{minted}{bash}
$: git clone https://gitlab.com/vcb-inclusive/kolya.git
\end{minted}
Afterwards, change the directory to the \texttt{kolya}
directory and install it with \texttt{pip3} in the following way:
\begin{minted}{bash}
$: cd kolya
$: pip3 install .
\end{minted} 
The dependencies will be automatically downloaded and installed during the setup. 
To get started, just import the package into a Python shell or a Jupyter notebook:
\begin{minted}{python}
>>> import kolya
\end{minted}
Note that the first time \texttt{kolya} is loaded, several functions are translated 
from Python to optimized machine code by Numba and cached. This stage may take from several seconds up to a few minutes.

\subsection{Parameter classes}
The library contains classes to store various real-valued variables.
One class is dedicated to the physical parameters like heavy quark masses and the strong coupling constant,
one for the HQE parameters, and one for the Wilson coefficients in the NP extension.
Dimensionful quantities, like the quark masses, are given in units of GeV.
The values of the physical parameters are stored in an object of \verb|parameters.physical_parameters| class
\begin{minted}{python}
>>> par = kolya.parameters.physical_parameters()
\end{minted}
The new object \texttt{par} contains information about $M_B$, $m_b^\mathrm{kin}(\mu_\mathrm{kin})$,
$\overline{m}_c(\mu_c)$ and $\alpha_s^{(4)}(\mu_s)$.
The bottom quark mass in the kinetic scheme and its scale are given in the variables \verb|mbkin| and \verb|scale_mbkin|,
while the charm mass in the $\overline{\mathrm{MS}}$ at renormalization scale $\mu_c$
corresponds to the variables \verb|mcMS| and \verb|scale_mcMS|.
The strong coupling constant $\alpha_s^{(4)}$ and its renormalization scale scale $\mu_s$ correspond to the 
class variables \verb|alphas| and \verb|scale_alphas|.
The class initializes also the renormalization scales of the HQE parameters $\mu_G^2,\ \rho_D^3$ and $\rho_{LS}^3$
through the variables \verb|scale_muG|, \verb|scale_rhoD| and \verb|scale_rhoLS|. At present, these are 
set equal to $m_b^\mathrm{kin}(\mu_\mathrm{kin})$.

The values stored in the object \verb|par|
can be shown with the \verb|show| method:
\begin{minted}{python}
>>> par.show()
bottom mass:       mbkin( 1.0  GeV)     =  4.563  GeV
charm mass:        mcMS( 3.0  GeV)      =  0.989  GeV
coupling constant: alpha_s( 4.563  GeV) =  0.2182
\end{minted}
where the current default values are based on the latest version of the FLAG 21 review~\cite{FlavourLatticeAveragingGroupFLAG:2021npn} as of February 2024~\cite{FLAG2024}.
Values different from the default ones can be set during initialization.
For instance, we can initialize the $\overline{m}_c(2\, \mathrm{GeV}) = 1.094$~GeV as follows:
\begin{minted}{python}
>>> par = kolya.parameters.physical_parameters(mcMS=1.094,scale_mcMS=2.0)
>>> par.show()
bottom mass:       mbkin( 1.0  GeV)     =  4.563  GeV
charm mass:        mcMS( 2.0  GeV)      =  1.094  GeV
coupling constant: alpha_s( 4.563  GeV) =  0.2182
\end{minted}
The example above shows that during initialization, the values of \verb|mcMS| and \verb|scale_mcMS|
must be set consistently. The following command
\begin{minted}{python}
>>> par = kolya.parameters.physical_parameters(mcMS=1.094)
\end{minted}
would initialize the charm mass to the (unphysical) value $\overline{m}_c(3\, \mathrm{GeV}) = 1.094$~GeV.
In order to set the quark masses at scales different from the default ones in a consistent way,
we include the method \verb|FLAG2024|. 
For instance, we set the quark masses at a scale $\mu_\mathrm{kin}=\mu_c=2$~GeV 
in the following way: 
\begin{minted}{python}
>>> par = kolya.parameters.physical_parameters()
>>> par.FLAG2024(scale_mcMS=2.0, scale_mbkin=2.0)
>>> par.show()
bottom mass:       mbkin( 2.0  GeV)     =  4.295730717092438  GeV
charm mass:        mcMS( 2.0  GeV)      =  1.0940623249384822  GeV
coupling constant: alpha_s( 4.563  GeV) =  0.21815198098622618
\end{minted}
Internally, the bottom and quark masses are recalculated using \texttt{CRunDec}~\cite{Chetyrkin:2000yt,Schmidt:2012az,Herren:2017osy} using the default values from Ref.~\cite{FLAG2024}. The scale of the strong coupling constant
can be modified in a similar way: 
\begin{minted}{python}
>>> par = kolya.parameters.physical_parameters()
>>> par.FLAG2024(scale_alphas=3.0)
>>> par.show()
bottom mass:       mbkin( 1.0  GeV)   =  4.56266484311551  GeV
charm mass:        mcMS( 3.0  GeV)    =  0.989  GeV
coupling constant: alpha_s( 3.0  GeV) =  0.2531150801276913
\end{minted}
Also in this case, we internally use \texttt{CRunDec} to evaluate $\alpha_s^{(4)}(3 \, \mathrm{GeV})$. 

The values of the HQE parameters in the historical basis (sometimes referred to as the ``perp'' basis in literature) are stored into an object of the class
\verb|parameters.HQE_parameters|. By default, their values are set to zero unless explicitly initialized:
\begin{minted}{python}
>>> hqe = kolya.parameters.HQE_parameters(
            muG = 0.306, 
            rhoD = 0.185, 
            rhoLS = -0.13,
            mupi = 0.477)
>>> hqe.show()
mupi  =  0.477  GeV^2
muG   =  0.306  GeV^2
rhoD  =  0.185  GeV^3
rhoLS =  -0.13  GeV^3
\end{minted}
where the values up to $1/m_b^3$ can be visualized all at once with \verb|show()|.
Since there are several operators at order $1/m_b^4$ and $1/m_b^5$,
we do not print them by default, however, their values can be inspected 
with the additional option \verb|show(flagmb4=1)| and \verb|show(flagmb5=1)|.
We introduce the class \verb|parameters.HQE_parameters_RPI| for the HQE parameters in the RPI basis:
\begin{minted}{python}
>>> hqe = kolya.parameters.HQE_parameters_RPI(
            muG = 0.306, 
            rhoD = 0.185, 
            mupi = 0.477)
\end{minted}
The parameter classes \verb|LSSA_HQE_parameters| and \verb|LSSA_HQE_parameters_RPI| contain numerical values for the HQE parameters in the historical and RPI basis, respectively, obtained using the ``lowest-lying state saturation ansatz'' (LLSA). The LLSA approximates the higher-order HQE parameters by expressing them through the bottom quark mass $m_b$, the HQE parameters $(\mu_\pi^2)^\perp$ and $(\mu_G^2)^\perp$, and the excitation energies $\epsilon_{1/2}$ and $\epsilon_{3/2}$. For further details on the LLSA, we refer to Ref.~\cite{Heinonen:2014dxa}, and for the expressions of the higher-order HQE parameters and their LLSA values, we refer to Refs.~\cite{Mannel:2023yqf, Mannel:2024crj}.

The Wilson coefficients of the effective dimension-six operators
defined in \eqref{eq::Hamiltonian-NP} are initialized via the class \verb|parameters.WCoefficients| and 
their values are inspected with the method \verb|show()|
\begin{minted}{python}
>>> wc = kolya.parameters.WCoefficients(SL=0.1,SR=-0.05)
>>> wc.show()
C_{V_L} =  0
C_{V_R} =  0
C_{S_L} =  0.1
C_{S_R} =  -0.05
C_{T} =  0
\end{minted}
The Wilson coefficients $C_{V_L}, C_{V_R}, C_{S_L}, C_{S_R}$ and $C_{T}$ are denoted 
by \verb|VL, VR, SL, SR| and \verb|T| respectively. By default, they are initialized to zero.

\subsection{Moment predictions}
We implemented the first four centralized moments of the $q^2$ spectrum and the
first three moments of $E_l$ and $M_X^2$. To evaluate them, we first need to 
initialize three objects for the physical parameters, the HQE parameters, and the 
Wilson coefficients:
\begin{minted}{python}
>>> par = kolya.parameters.physical_parameters()
>>> hqe = kolya.parameters.HQE_parameters(
            muG = 0.306, 
            rhoD = 0.185, 
            rhoLS = -0.13,
            mupi = 0.477)
>>> wc = kolya.parameters.WCoefficients()
\end{minted}
The prediction for the $q^2$ moments receive as 
inputs the value of $q^2_\mathrm{cut}$ expressed in GeV$^2$,
and the three objects \verb|par, hqe| and \verb|wc|.
The value of each $q^2$ moment is obtained with the functions
\verb|Q2moments.moment_n_KIN_MS(q2cut, par, hqe, wc)|, where \verb|n|$=1,2,3,4$.
For example, to evaluate $q_1(q^2_\mathrm{cut})$ for  $q^2_\mathrm{cut} = 8.0$~GeV$^2$, type
\begin{minted}{python}
>>> q2cut = 8.0   #GeV^2
>>> kolya.Q2moments.moment_1_KIN_MS(q2cut, par, hqe, wc)
8.996406491856465 #GeV^2
\end{minted}
The result is provided in the respective powers of GeV$^{2n}$. 
The suffix \texttt{KIN} and \texttt{MS} refers to the scheme
for bottom (kinetic) and charm ($\overline{\mathrm{MS}}$) masses.
By default, the evaluation considers
power corrections up to $1/m_b^3$. The corrections of order $1/m_b^4$ and $1/m_b^5$ 
can be included by setting the optional arguments \verb|flagmb4=1| and \verb|flagmb5=1|.
For instance, we can set the higher-order HQE parameters $m_1=0.1 \, \text{GeV}^4$ and $r_1=0.1 \, \text{GeV}^5$
and compare the predictions up to order $1/m_b^5$ in the following way:
\begin{minted}{python}
>>> hqe.m1=0.1 #GeV^4
>>> hqe.r1=0.1 #GeV^5
>>> kolya.Q2moments.moment_1_KIN_MS(q2cut, par, hqe, wc)
8.996406491856465
>>> kolya.Q2moments.moment_1_KIN_MS(q2cut, par, hqe, wc, flagmb4=1)
8.966491804532277
>>> kolya.Q2moments.moment_1_KIN_MS(q2cut, par, hqe, wc, flagmb4=1, flagmb5=1)
8.75600814954277
\end{minted}
Setting \verb|flagmb4=0| and \verb|flagmb5=0| eliminates all terms of order $1/m_b^4$ and $1/m_b^5$, respectively. We note that at these orders also mixing terms proportional to $\mu_G^2 \mu_\pi^2$ or $\mu_G^2 \rho_D^3$ enter which can only be excluded by putting these two flags to zero. Therefore, putting these flags to zero does not have the same effect as simply setting all the $1/m_b^4$ and $1/m_b^5$ HQE element to zero. 

Concerning the perturbative corrections, these are all included by default in the 
numerical evaluation (see Tab.~\ref{tab:overview} for the current orders in $\alpha_s$ implemented). 
For cross-checks with the literature and the study of their impact, the NNLO corrections 
can be switched off via the optional argument \verb|flag_includeNNLO=0| (the default is \verb|flag_includeNNLO=1|).
Also, the NLO corrections to the power-suppressed terms can be excluded with \verb|flag_includeNLOpw=0|.

Moreover, the option \verb|flag_DEBUG=1| will print a report of the various contributions coming from
the higher-order QCD corrections:
\begin{minted}{python}
>>> kolya.Q2moments.moment_1_KIN_MS(8.0, par, hqe, wc, flag_DEBUG=1)
Q2moment n. 1 LO =  9.148659808170105
Q2moment n. 1 NLO = api * -1.319532010835962
Q2moment n. 1 NNLO = api^2 * -9.616956902561078
Q2moment n. 1 NLO pw = api * -0.7873907726673756
Q2moment n. 1 NNLO from NLO pw = api^2 * 8.39048437244325
\end{minted}
The contributions denoted by \verb|NLO| and \verb|NNLO| are the coefficients in front of
$\alpha_s(\mu_s)/\pi$ and $(\alpha_s(\mu_s)/\pi)^2$ to leading order in $1/m_b$.
The term \verb|NLO pw| corresponds to the overall NLO correction in the terms of 
order $1/m_b^2$ and $1/m_b^3$. In the kinetic scheme, the inclusion of the NLO corrections 
to the power-suppressed terms induces also an additional $O(\alpha_s^2)$ contribution
to leading order in $1/m_b$, which is reported in the last line.

The predictions for the $E_l$ and $M_X^2$ moments follow a similar syntax. 
The first argument passed to the function corresponds to the value of the cut $E_\mathrm{cut}$ in units of GeV. 
For instance, the first moments of $E_l$ and $M_X^2$ for $E_\mathrm{cut}=1.0$~GeV
are evaluated as follows:
\begin{minted}{python}
>>> par = kolya.parameters.physical_parameters()
>>> hqe = kolya.parameters.HQE_parameters(
            muG = 0.306, 
            rhoD = 0.185, 
            rhoLS = -0.13,
            mupi = 0.477)
>>> wc = kolya.parameters.WCoefficients()
>>> elcut=1.0 #GeV
>>> kolya.Elmoments.moment_1_KIN_MS(elcut, par, hqe, wc)
1.549535385165418 #GeV
>>> kolya.MXmoments.moment_1_KIN_MS(elcut, par, hqe, wc)
4.3850057885330544 #GeV^2
\end{minted}
Higher moments $\ell_2,\ell_3, h_2, h_3$ are computed in a similar way by replacing 
\verb|moment_1| with \verb|moment_2| or \verb|moment_3|. The result for $\ell_n(E_\mathrm{cut})$ is in GeV$^n$,
while for $h_n(E_\mathrm{cut})$ the result is in GeV$^{2n}$.

By default, the moments are calculated using the HQE parameters as defined in the historical basis. 
The $q^2$ moments and the total rate can also be calculated using the RPI basis adopted in Ref.~\cite{Bernlochner:2022ucr}.
The predictions in the RPI basis are obtained by passing the optional argument \verb|flag_basisPERP=0|. In this case, 
the HQE parameters must be passed to the function through an object of the class \verb|HQE_parameters_RPI|:
\begin{minted}{python}
>>> par = kolya.parameters.physical_parameters()
>>> hqeRPI = kolya.parameters.HQE_parameters_RPI(
            muG = 0.38, 
            rhoD = 0.03, 
            mupi = 0.43)
>>> wc = kolya.parameters.WCoefficients()
>>> kolya.Q2moments.moment_1_KIN_MS(8.0, par, hqeRPI, wc, flag_basisPERP=0)
9.350141389366614
\end{minted}
The RPI basis is supported only for $q_n$ moments and the total rate since reparametrization invariance 
reduces the number of HQE parameters only for them. Similar to before, the $1/m_b^4$ and $1/m_b^5$ corrections in the RPI basis can be included by using the optional arguments \verb|flagmb4=1| and \verb|flagmb5=1|.
For the $\ell_n$ and $h_n$, we stick to the historical basis.

\subsection{Branching ratio prediction}
To obtain the branching ratio $\mathrm{Br}(B \to X_c l \bar \nu_l)$ or the total semileptonic width $\Gamma_\mathrm{sl}$,
three objects for the physical parameters, the HQE parameters, and the Wilson coefficients must be 
initialized, as discussed for the moments in the previous subsection.
For the total rate $\Gamma_\mathrm{sl}$ defined in \eqref{eqn:GammaSL}, type 
\begin{minted}{python}
>>> Vcb = 42.2e-3
>>> kolya.TotalRate.TotalRate_KIN_MS(Vcb, par, hqe, wc)
4.4016320941077224e-14 #GeV
\end{minted}
where the first argument is the value of $|V_{cb}|$ and the result is expressed in GeV.
For the evaluation of the branching ratio, we use 
\begin{minted}{python}
>>> Vcb = 42.2e-3
>>> kolya.TotalRate.BranchingRatio_KIN_MS(Vcb, par, hqe, wc)
0.10555834162102022
\end{minted}
We obtain the branching ratio by dividing $\Gamma_\mathrm{sl}$ by the average lifetime
of the $B^\pm$ and $B_0$ mesons. 

The partial width $\Delta \Gamma_\mathrm{sl}(E_\mathrm{cut})$ 
with cut on $E_l$ is obtained with 
\begin{minted}{python}
>>> Vcb = 42.2e-3
>>> Elcut = 1.0 # GeV
>>> kolya.DeltaBR.DeltaRate_KIN_MS(Vcb, Elcut, par, hqe, wc)
3.469103884950744e-14 #GeV
\end{minted}
where the first argument is the value of $|V_{cb}|$, the second argument 
the value of $E_\mathrm{cut}$ in GeV. The result is reported in GeV.
The corresponding value for the branching ratio is given by
\begin{minted}{python}
>>> Vcb = 42.2e-3
>>> Elcut = 1.0 # GeV
>>> kolya.DeltaBR.DeltaBR_KIN_MS(Vcb, Elcut, par, hqe, wc)
0.08319478892764474
\end{minted}

The function that predicts the branching ratio allows optional arguments \verb|flagmb4=1| and \verb|flagmb5=1|
to include the power corrections of order $1/m_b^4$ and $1/m_b^5$. The predictions in the RPI basis are obtained by passing the optional argument \verb|flag_basisPERP=0|.
For cross-checks with the literature and the study of the impact of QCD corrections, 
the NNLO and N3LO corrections to the total rate can be switched off via the optional arguments
\verb|flag_includeNNLO=0| and \verb|flag_includeN3LO=0| (by default, all these corrections are included). 
Moreover, the effects arising from the NLO corrections to the power-suppressed terms 
can be excluded with \verb|flag_includeNLOpw=0|.

\section{Outlook \& Conclusion}
\label{sec:conc}

In this document, we have presented the first version of the open-source library \texttt{kolya}, corresponding to the release 1.0. 
In this release, we have implemented the predictions in the HQE for the total rate and the moments of $q^2$, $E_l$ and $M_X^2$. Currently, this is sufficient for comparison with published experimental results by $B$ factories.
We included all higher order corrections in $\alpha_s$ and $1/m_b$ which are available at this specific point in time
and are summarized in Tab.~\ref{tab:overview}.

On the GitLab repository, we provide, additionally, interactive tutorials running as a Jupyter notebook and validation notebooks which demonstrate how 
the library can reproduce the results available in the 
literature.
The library is open source, so code contributions and improvements are very welcome. 
In particular, new higher-order corrections can be implemented like
\begin{itemize}
    \item QED effects calculated in Ref.~\cite{Bigi:2023cbv},
    \item exact results for the
    NNLO corrections to $E_l$ and $M_X^2$ moments with a
    lower cut $E_\mathrm{cut}$,
    \item renormalization group evolution 
    of the HQE parameters to NLO,
    \item the NLO corrections in the 
    coefficients of $\rho_D^3$ and $\rho_{LS}^3$ for 
    the $E_l$ and $M_X^2$ moments.
\end{itemize}
Additional observables can play an important role in 
better improving the extraction of the HQE parameters
or have an important role in testing the SM.
These new observables may include
\begin{itemize}
    \item forward-backward asymmetries $A_{FB}$
    and $q^2$, $E_l$ and $M_X^2$ moments for forward and backward events~\cite{Turczyk:2016kjf,Herren:2022spb},
    \item the ratio $R_X = \Gamma_{B \to X_c \tau \bar \nu_\tau}/\Gamma_{B \to X_c l \bar \nu_l} $,
    \item the lifetime of $B$ mesons within the HQE,
    \item predictions for the decay into charmless 
    final states    $B \to X_u l\bar \nu_l$.
\end{itemize}
Finally, \texttt{kolya} could also be extended to include predictions for inclusive $D$ decays discussed in detail in \cite{Fael:2019umf}.

\section*{ Acknowledgments } 
We thank F.\ Bernlochner and M.\ Prim for ongoing collaboration and
suggestions and P.\ Gambino for providing the results of Refs.~\cite{Aquila:2005hq,Alberti:2012dn,Alberti:2013kxa} 
in electronic form.
We also thank D.\ Straub for discussion about the Python package 
\texttt{python-rundec}~\cite{pythonrundec}, which provides a wrapper around \texttt{CRunDec}~\cite{Chetyrkin:2000yt,Schmidt:2012az,Herren:2017osy}.
The work of M.F.\ is supported by the European Union’s Horizon 2020 research and innovation
program under the Marie Sk\l{}odowska-Curie grant agreement No.~101065445 - PHOBIDE. The work of I.S.M. was supported by the Deutsche Forschungsgemeinschaft (DFG, German Research Foundation) under grant 396021762 -- TRR 257 ``Particle Physics Phenomenology after the Higgs Discovery". K.K.V. acknowledges support from the project “Beauty decays: the quest for extreme
precision” of the Open Competition Domain Science which is financed by the Dutch Research
Council (NWO).

\appendix
\section{Definition of the HQE elements}\label{app:HQE_def}
Here, we define the HQE parameters both in the historical and the RPI basis up to $1/m_b^5$. The conversions between these two bases can be found in Ref.~\cite{Mannel:2023yqf}. 

\subsection{Historical basis}
The HQE matrix elements in the historical basis, denoted by ``$\perp$'', are defined through the spacial covariant derivatives $iD^\perp_\mu=g^\perp_{\mu\nu}iD^\nu$, where 
\begin{align}
    g^\perp_{\mu\nu}&=g_{\mu\nu}-v_\mu v_\nu\ ,
\end{align} as in Refs.~\cite{Mannel:2010wj, Gambino:2016jkc}. We will employ here the notation $\langle \bar{b}_v\,...\,b_v\rangle\equiv\langle B(v)|\bar{b}_v\,...\,\bar{b}_v|B(v)\rangle$. At $1/m_b^2$, we have
\begin{align}
    2m_B(\mu_\pi^2)^\perp&=-\langle \bar{b}_v\, (iD^\rho)\, (iD^\sigma)\, b_v\rangle\, g_{\rho\sigma}^\perp\ ,\nonumber\\
    2m_B(\mu_G^2)^\perp&=\frac{1}{2}\langle\bar{b}_v \, \big[(iD^\rho),\, (iD^\sigma)\big]\, (-i\sigma^{\alpha\beta})\, b_v\rangle\, g_{\rho\alpha}^\perp g_{\sigma\beta}^\perp\ ,
\end{align}
where $\gamma^\mu\gamma^\nu=g^{\mu\nu}+(-i\sigma^{\mu\nu})$. At $1/m_b^3$, we have
\begin{align}
    2m_B(\rho_D^3)^\perp&= \frac{1}{2}\langle \bar{b}_v\, \Big[(iD^\rho),\, \big[(iD^\sigma),\, (iD^\lambda)\big]\Big]\,b_v\rangle\, g^\perp_{\rho\lambda} v_\sigma\ ,\nonumber\\
    2m_B(\rho_{LS}^3)^\perp&=\frac{1}{2}\langle \bar{b}_v\, \Big\{(iD^\rho),\, \big[(iD^\sigma),\, (iD^\lambda)\big]\Big\}\, (-i\sigma^{\alpha\beta})\,b_v\rangle\, g^\perp_{\rho\alpha}g^\perp_{\lambda\beta}v_\sigma\ .
\end{align}
The nine HQE parameters at $1/m_b^4$ were first introduced in Ref.~\cite{Mannel:2010wj}. We list them here:
\begin{align}
    2m_B m_1&=\langle \bar{b}_v\, (iD^\rho)\,(iD^\sigma)\,(iD^\lambda)\,(iD^\delta) \,b_v\rangle\, \frac{1}{3}\left(g^\perp_{\rho\sigma}g^\perp_{\lambda\delta}+g^\perp_{\rho\lambda}g^\perp_{\sigma\delta}+g^\perp_{\rho\delta}g^\perp_{\sigma\lambda}\right)\ ,\nonumber\\
    2m_B m_2&=\langle \bar{b}_v\, \big[(iD^\rho),\, (iD^\sigma)\big]\,\big[(iD^\lambda),\, (iD^\delta)\big]\,b_v\rangle\, g^\perp_{\rho\delta}v_\sigma v_\lambda\ ,\nonumber\\
    2m_B m_3&=\langle \bar{b}_v\, \big[(iD^\rho),\, (iD^\sigma)\big]\,\big[(iD^\lambda),\, (iD^\delta)\big]\,b_v\rangle\, g^\perp_{\rho\lambda}g^\perp_{\sigma\delta}\ ,\nonumber\\
    2m_B m_4&=\langle \bar{b}_v\, \Big\{(iD^\rho),\, \Big[(iD^\sigma),\, \big[(iD^\lambda),\, (iD^\delta)\big]\Big]\Big\}\,b_v\rangle\, g^\perp_{\sigma\lambda}g^\perp_{\rho\delta}\ ,\nonumber \rule[-10pt]{0pt}{8pt}\\  
    2m_B m_5&=\langle \bar{b}_v\, \big[(iD^\rho),\, (iD^\sigma)\big]\,\big[(iD^\lambda),\, (iD^\delta)\big]\, (-i\sigma^{\alpha\beta})\,b_v\rangle\, g^\perp_{\alpha\rho}g^\perp_{\beta\delta}v_\sigma v_\lambda\ ,\nonumber\\
    2m_B m_6&=\langle \bar{b}_v\, \big[(iD^\rho),\, (iD^\sigma)\big]\,\big[(iD^\lambda),\, (iD^\delta)\big]\, (-i\sigma^{\alpha\beta})\,b_v\rangle\, g^\perp_{\alpha\sigma}g^\perp_{\beta\lambda}g^\perp_{\rho\delta}\ ,\nonumber\\
    2m_B m_7&=\langle \bar{b}_v\, \Big\{\big\{(iD^\rho),\, (iD^\sigma)\big\},\,\big[(iD^\lambda),\, (iD^\delta)\big]\Big\}\, (-i\sigma^{\alpha\beta})\,b_v\rangle\, g^\perp_{\sigma\lambda}g^\perp_{\alpha\rho}g^\perp_{\beta\delta}\ ,\nonumber\\
    2m_B m_8&=\langle \bar{b}_v\,\Big\{\big\{(iD^\rho),\, (iD^\sigma)\big\},\,\big[(iD^\lambda),\, (iD^\delta)\big]\Big\}\, (-i\sigma^{\alpha\beta}) \,b_v\rangle\, g^\perp_{\rho\sigma}g^\perp_{\alpha\lambda}g^\perp_{\beta\delta}\ ,\nonumber\\
    2m_B m_9&=\langle \bar{b}_v\, \Bigg[(iD^\rho),\, \Big[(iD^\sigma),\, \big[(iD^\lambda),\, (iD^\delta)\big]\Big]\Bigg]\, (-i\sigma^{\alpha\beta})\,b_v\rangle\, g^\perp_{\rho\beta}g^\perp_{\lambda\alpha}g^\perp_{\sigma\delta}\ .
\end{align}
Finally, eighteen more parameters are present at $1/m_b^5$, as defined in Ref.~\cite{Mannel:2010wj}:
\begin{align}
\nonumber
2m_B r_1 &= \langle \bar{b}_v \,(i  D_\mu)\, (ivD)^3\, (i  D^\mu) \, b_v\rangle\ , \\
\nonumber 
2m_B r_2 &= \langle \bar{b}_v \,(i  D_\mu)\, (ivD)\, (i  D^\mu)\, (iD)^2 \, b_v\rangle\ , \\
\nonumber
2m_B r_3 &= \langle \bar{b}_v \,(i  D_\mu)\, (ivD)\,(i  D_\nu)\,( i
D^\mu)\,(i  D^\nu )\, b_v\rangle\ , \\  
\nonumber
2m_B r_4 &= \langle \bar{b}_v \,(i  D_\mu)\, (ivD)\, (iD)^2\,(i  D^\mu )\, b_v\rangle\ , \\  
\nonumber
2m_B r_5 &= \langle \bar{b}_v \,(iD)^2\,(ivD)\,  (iD)^2 \, b_v\rangle\ , \\  
\nonumber
2m_B r_6 &= \langle \bar{b}_v \,(i  D_\mu)\,(i  D_\nu)\, (ivD)\, (i
D^\nu)\,(i  D^\mu )\, b_v\rangle\ , \\  
\nonumber
2m_B r_7 &= \langle \bar{b}_v \,(i  D_\mu)\,(i  D_\nu)\, (ivD)\, (i
D^\mu)\,(i  D^\nu) \, b_v\rangle\ , \rule[-10pt]{0pt}{8pt}\\  
\nonumber
2m_B r_{8} &= \langle \bar{b}_v \,(i   D_\alpha) \, (ivD)^3\,( i   D_\beta)
\, (-i \sigma^{\alpha \beta })\,b_v\rangle\ , \\  
\nonumber
2m_B r_{9} &= \langle \bar{b}_v \,(i   D_\alpha) \, (ivD)\,( i   D_\beta)
\, (iD)^2 \,(-i \sigma^{\alpha \beta })\, b_v\rangle\ , \\  
\nonumber
2m_B r_{10} &= \langle \bar{b}_v \,(i   D_\mu)\, (ivD)\, (i
D^\mu)\, (i   D_\alpha) \,( i   D_\beta)  \,(-i \sigma^{\alpha \beta })\, b_v\rangle\ , \\
\nonumber
2m_B r_{11} &= \langle \bar{b}_v \,(i   D_\mu)\, (ivD)\, (i   D_\alpha)
\, (i   D^\mu)\,( i   D_\beta)  \,(-i \sigma^{\alpha \beta })\, b_v\rangle\ , \\  
\nonumber
2m_B r_{12} &= \langle \bar{b}_v \,(i   D_\alpha) \, (ivD)\, (i
D_\mu)\,( i   D_\beta) \, (i   D^\mu )\,(-i \sigma^{\alpha \beta })\, b_v\rangle\ , \\
\nonumber
2m_B r_{13} &= \langle \bar{b}_v \,(i   D_\mu)\, (ivD)\, (i   D_\alpha)
\,( i   D_\beta) \, (i   D^\mu )\,(-i \sigma^{\alpha \beta })\, b_v\rangle\ , \\  
\nonumber
2m_B r_{14} &= \langle \bar{b}_v \,(i   D_\alpha) \, (ivD)\, (iD)^2\,( i   D_\beta)  \,(-i \sigma^{\alpha \beta })\, b_v\rangle\ , \\
\nonumber
2m_B r_{15} &= \langle \bar{b}_v \,(i   D_\alpha) \,( i   D_\beta) \, (ivD)\, (iD)^2 \,(-i \sigma^{\alpha \beta })\, b_v\rangle\ , \\  
\nonumber
2m_B r_{16} &= \langle \bar{b}_v \,(i   D_\mu)\,( i   D_\alpha) \, (ivD)\,( i   D_\beta) \, (i   D^\mu )\,(-i \sigma^{\alpha \beta })\, b_v\rangle\ , \\  
\nonumber
2m_B r_{17} &= \langle \bar{b}_v \,(i   D_\alpha) \,( i   D_\mu)\, (ivD)\, (i   D^\mu)\,( i   D_\beta)  \,(-i \sigma^{\alpha \beta })\, b_v\rangle\ , \\  
2m_B r_{18} &= \langle \bar{b}_v \,(i   D_\mu)\, (i   D_\alpha) \, (ivD)\, (i   D^\mu)\,( i   D_\beta)  \,(-i \sigma^{\alpha \beta })\, b_v\rangle \ .\label{ri_definitions}
\end{align}
\subsection{RPI basis}
The RPI HQE matrix elements up to $1/m_b^4$ have been determined in Ref.~\cite{Mannel:2018mqv}. We list them here:
\begin{align}
    2m_B\mu_\pi^2&=-\langle \bar{b}_v\, (iD)^2\, b_v\rangle\ ,\nonumber \\
    2m_B\mu_G^2&=\langle \bar{b}_v\, (iD_\alpha)\,(iD_\beta)\,(-i\sigma^{\alpha\beta})\,b_v\rangle\ ,\nonumber\\
    2m_B\tilde{\rho}_D^3&=\frac{1}{2}\langle \bar{b}_v\,\Bigg[(iD_\mu),\Big[\Big((ivD)+\frac{1}{2m_b}(iD)^2\Big),(iD^\mu)\Big]\Bigg]\,b_v\rangle\ ,\nonumber\\
    2m_Br_G^4&=\langle \bar{b}_v\,\big[(iD_\mu),\,(iD_\nu)\big]\,\big[(iD^\mu),\,(iD^\nu)]\,b_v\rangle\ ,\nonumber\\
    2m_B\tilde{r}_E^4&=\langle \bar{b}_v\,\Big[\Big((ivD)+\frac{1}{2m_b}(iD)^2\Big),\,(iD_\mu)\Big]\,\Big[\Big((ivD)+\frac{1}{2m_b}(iD)^2\Big),\,(iD^\mu)\Big]\,b_v\rangle\ ,\nonumber\\
    2m_Bs_B^4&=\langle \bar{b}_v\,\big[(iD_\mu),\,(iD_\alpha)\big]\,\big[(iD^\mu),\,(iD_\beta)\big]\,(-i\sigma^{\alpha\beta})\,b_v\rangle\ ,\nonumber\\
    2m_B\tilde{s}_E^4&=\langle \bar{b}_v\,\Big[\Big((ivD)+\frac{1}{2m_b}(iD)^2\Big),\,(iD_\alpha)\Big]\,\Big[\Big((ivD)+\frac{1}{2m_b}(iD)^2\Big),\,(iD_\beta)\Big]\,(-i\sigma^{\alpha\beta})\,b_v\rangle ,\nonumber\\
    2m_Bs_{qB}^4&=\langle\bar{b}_v\,\Bigg[(iD_\mu),\,\Big[(iD^\mu),\,\big[(iD_\alpha),\,(iD_\beta)\big]\Big]\Bigg]\,(-i\sigma^{\alpha\beta})\,b_v\rangle\ ,
\end{align}
We note that we choose to use $\mu_\pi^2$ instead of $\mu_3$ as in Ref.~\cite{Mannel:2018mqv}, in order to avoid factors of $1/\mu_3$ in the centralized moments. Furthermore, we note that $\tilde{\rho}_D^3$, $\tilde{r}_E^4$, and $\tilde{s}_E^4$ contain their so-called RPI-completion terms as described in Refs.~\cite{Mannel:2018mqv, Mannel:2023yqf}. In Ref.~\cite{Mannel:2023yqf}, the RPI matrix elements at $1/m_b^5$ have been determined to be:
\begin{align}
    2m_BX_1^5&=\langle \bar{b}_v\,\Big[(ivD),\,\big[(ivD),\,(iD_\mu)\big]\Big]\,\big[(ivD),\,(iD^\mu)\big]\,b_v\rangle\ \nonumber,\\
      2m_BX_2^5&=\langle \bar{b}_v\,\Big[ (ivD),\,\big[(iD_\mu),\,(iD_\nu)\big]\Big]\,\big[(iD^\mu),\,(iD^\nu)\big]\,b_v\rangle\ , \nonumber\\
      2m_BX_3^5&= \langle \bar{b}_v \,\Big[(iD_\mu),\,\big[(ivD),\,(iD_\nu)\big]\,\big[(iD^\mu),\,(iD^\nu)\big]\Big]\,b_v\rangle\ 
 \nonumber\\
     2m_BX_4^5 &=\langle \bar{b}_v\, \Bigg[(iD_\mu),\,\Bigg[(iD_\nu),\,\Big[(iD^\mu),\,\big[(ivD),\,(iD^\nu)\big]\Big]\Bigg]\Bigg]\,b_v\rangle\ , \nonumber\rule[-10pt]{0pt}{8pt}\\ 
       2m_BX_5^5&=\langle \bar{b}_v\,\Big[(ivD),\,\big[(ivD),\,(iD_\alpha)\big]\Big]\,\big[(ivD),\,(iD_\beta)\big]\,(-i\sigma^{\alpha\beta})\,b_v\rangle\ ,\nonumber\\
       2m_BX_6^5&=\langle \bar{b}_v\,\Big[(ivD),\,\big[(iD_\mu),\,(iD_\alpha)\big]\Big]\,\big[(iD^\mu),\,(iD_\beta)]\,(-i\sigma^{\alpha\beta})\,b_v\rangle\ ,\nonumber \\
       2m_BX_7^5&=\langle \bar{b}_v\,\Big[(iD_\mu),\,\big[(ivD),\,(iD_\alpha)\big]\Big]\,\big[(iD^\mu),\,(iD_\beta)\big]\,(-i\sigma^{\alpha\beta})\,b_v\rangle\ , \nonumber\\
    2m_BX_8^5&=\langle \bar{b}_v\,\Big[(iD_\mu),\,\big[(ivD),\,(iD_\alpha)\big]\,\big[(iD^\mu),\,(iD_\beta)\big]\Big]\,(-i\sigma^{\alpha\beta})\,b_v\rangle\ ,  \nonumber\\
    2m_BX_9^5&=\langle \bar{b}_v\,\Big[(iD_\mu),\,\big[(ivD),\,(iD^\mu)\big]\Big]\,\big[(iD_\alpha),\,(iD_\beta)\big]\,(-i\sigma^{\alpha\beta})\,b_v\rangle\ , \nonumber\\
    2m_BX_{10}^5&=\langle\bar{b}_v\,\Big[(iD_\mu),\,\big[(ivD),\,(iD^\mu)\big]\,\big[(iD_\alpha),\,(iD_\beta)\big]\Big]\,(-i\sigma^{\alpha\beta})\,b_v\rangle\ .
\end{align}

\newpage
\bibliographystyle{JHEP.bst}
\bibliography{refs.bib,non-inspires-refs.bib}

\end{document}